\documentclass[useAMS,usenatbib]{mn2e_mod}
\usepackage{natbib}
\usepackage{graphicx}
\usepackage{amsmath}
\usepackage{amssymb}
\usepackage{deluxetable}
\usepackage{longtable}
\usepackage{url}

\let\ACMmaketitle=\maketitle
\renewcommand{\maketitle}{\begingroup\let\footnote=\thanks \ACMmaketitle\endgroup}

\title[multi-phase outflows in SDSS J124754.95-033738.6]{Multi-phase outflows in post starburst E+A galaxies - II. A direct connection between the neutral and ionized outflow phases}

\author[Baron et al.]
{Dalya Baron$^{1}$\thanks{dalyabaron@gmail.com},
Hagai Netzer$^{1}$,
Ric I. Davies$^{2}$ \& 
J. Xavier Prochaska$^{3}$
\\
\\
$^{1}$School of Physics and Astronomy, Tel-Aviv University, Tel Aviv 69978, Israel.\\
$^{2}$Max-Planck-Institut f$\mathrm{\ddot{u}}$r extraterrestrische Physik, Postfach 1312, D-85741, Garching, Germany. \\
$^{3}$Department of Astronomy and Astrophysics, UCO/Lick Observatory, University of California, 1156 High Street, Santa Cruz, CA 95064, USA.
}

\begin{document}

\maketitle

\label{firstpage}
\begin{abstract}

Post starburst E+A galaxies are systems that hosted a powerful starburst that was quenched abruptly. Simulations suggest that these systems provide the missing link between major merger ULIRGs and red and dead ellipticals, where AGN feedback is responsible for the expulsion or destruction of the molecular gas. However, many details remain unresolved and little is known about AGN-driven winds in this short-lived phase. We present spatially-resolved IFU spectroscopy with MUSE/VLT of SDSS J124754.95-033738.6, a post starburst E+A galaxy with a recent starburst that started 70 Myrs ago and ended 30 Myrs ago, with a peak SFR of $\sim 150\,\mathrm{M_{\odot}/yr}$. We detect disturbed gas throughout the entire field of view, suggesting triggering by a minor merger. We detect fast-moving multi-phased gas clouds, embedded in a double-cone face-on outflow, which are traced by ionized emission lines and neutral NaID \emph{emission} and absorption lines. We find remarkable similarities between the kinematics, spatial extents, and line luminosities of the ionized and neutral gas phases, and propose a model in which they are part of the same outflowing clouds, which are exposed to both stellar and AGN radiation. Our photoionization model provides consistent ionized line ratios, NaID absorption optical depths and EWs, and dust reddening. Using the model, we estimate, for the first time, the neutral-to-ionized gas mass ratio (about 20), the sodium neutral fraction, and the size of the outflowing clouds. This is one of the best ever observed direct connections between the neutral and ionized outflow phases in AGN. 

\end{abstract}


\vspace{1cm}
\section{Introduction}\label{s:intro}

The discovery of strong correlations between the masses of super massive black holes (SMBHs) and several properties of their host galaxy (stellar velocity dispersion, bulge mass, and bulge luminosity; e.g., \citealt{gebhardt00a, ferrarese00, tremaine02, gultekin09}) have led to various suggestions about the connection between SMBH growth and stellar mass growth in galaxies (e.g., \citealt{silk98, kauffmann00, zubovas14}). Active galactic nuclei (AGN) feedback, in the form of powerful outflows, had been suggested as a way to couple the energy released by the accreting SMBH with the ISM of its host galaxy, providing the necessary link between the SMBH and its host galaxy growth. In particular, several galaxy formation theories suggest that during the AGN phase, the AGN drives galactic-scale winds that expel gas from its host galaxy, shuts down additional gas accretion onto the SMBH, terminates the star formation (SF) in the galaxy, and enriches its circumgalactic medium (CGM) with metals \citep{silk98, fabian99, benson03, king03, dimatteo05, hopkins06, gaspari11}. However, more recent zoom-in simulations suggest that these winds escape along the path of least resistance, with little impact on the bulk of gas in the galaxy (e.g., \citealt{gabor14, hartwig18, nelson19}).

Galactic scale winds are ubiquitous and span a large range of host galaxy properties (see recent review by \citealt{veilleux20}). They are detected is systems at different evolutionary stages, from star-bursting ultra luminous infrared galaxies (ULIRGs), through typical main sequence galaxies, to quenched elliptical galaxies (e.g., \citealt{rupke05b, rupke05c, mullaney13, veilleux13, cazzoli16, cheung16, fiore17, rupke17, forster_schreiber18, baron19a, baron19b}). They are detected on different physical scales, from the vicinity of the SMBH (e.g., \citealt{blustin03, reeves03, tombesi10}), to galactic scales, at hundreds of parsecs (\citealt{husemann16, fischer18, tadhunter18, baron19a}) to 1-10 kpc (\citealt{cano12, liu13a, harrison14, rupke17, leung19}). They are detected through different gas phases, from high velocity X-ray and UV absorption lines (e.g. \citealt{blustin03, reeves03, tombesi10, arav13}), through ionized emission lines \citep{mullaney13, harrison14, perna17, rupke17, baron19b, mingozzi19, rojas19, shimizu19}, to atomic and molecular emission and absorption lines \citep{feruglio10, veilleux13, cicone14, burillo15, cazzoli16, rupke17}.

The nature of multi-phased (molecular, atomic, and ionized gas) outflows remains largely unconstrained, and different phases \emph{detected in different systems} show different outflow velocities, covering factors, and mass (e.g., \citealt{fiore17, cicone18, veilleux20}). It is unclear whether these phases are connected, and multi-phased outflow studies were only conducted in a handful of sources so far (see comprehensive review by \citealt{cicone18}). For example, IC5063 is the only galaxy where the ionized, neutral, and molecular phases of the outflow show similar kinematics and spatial extents \citep{tadhunter14}. \citet{rupke05c} and \citet{rupke17} conducted a detailed comparison between the neutral and ionized outflow phases in their sample, finding similarities in some of the systems. Using the infrared emission of dust that is mixed with the ionized outflow (\citealt{baron19a}), we suggested the existence of a significant amount of neutral atomic gas at the back of the outflowing ionized gas clouds in a large sample of type II AGN (\citealt{baron19b}), which has significant implications for the estimated mass and energetics of such flows.

There are two major uncertainties concerning AGN-driven feedback in active galaxies. The first is related to the relative contribution of AGN versus supernovae to the observed winds, where the latter is directly related to SF activity in the galaxy. This is due to the fact that most systems showing AGN activity also show significant SF activity, with some correlation between the AGN bolometric luminosity and the SF luminosity (e.g. \citealt{netzer09}). Many systems with more powerful AGN, that are capable of producing stronger AGN-driven winds, also undergo powerful SF episodes, capable of driving stronger supernovae-driven winds. Studies typically assume that the source that photoionizes the stationary and outflowing gas is also the main driver of the observed winds (outflows that are observed in systems with emission line ratios that are consistent with AGN photoionization, are considered to be driven by the AGN; e.g., \citealt{mullaney13, karouzos16b, forster_schreiber18, leung19}). However, in \citet{baron19b}, using a large sample of type II AGN, we found evidence that although the AGN dominates the ionization of the outflowing gas, the ionized outflows are more likely to be driven by supernovae (see also \citealt{husemann19}). 

The second uncertainty is related to the timescale of the observed winds. To quantify the effect of these outflows on the host galaxy, it is important to map the evolutionary stages in which the feedback is most prominent. In addition, the total energy that is transferred from the accreting SMBH to its host galaxy ISM depends on the mass outflow rate and on the duration of these flows. Unfortunately, in the large majority of systems where outflows are currently detected, it is difficult to estimate the total duration of the feedback episode (see however example in \citealt{baron18}). 

Post starburst E+A galaxies offer an advantage over other galaxy samples in dealing with these uncertainties. Their optical spectra show a narrow stellar age distribution, with a significant contribution from A-type stars, and no contribution from O and B stars, indicating a recent starburst that was quenched abruptly (e.g., \citealt{dressler99, poggianti99, goto04, dressler04, french18}). The estimated SFRs during the burst range from 50 to 300 $\mathrm{M_{\odot}/yr}$ \citep{poggianti00, kaviraj07}, and the mass fractions forming in the burst are high, 10\%--80\% of the total stellar mass \citep{liu96, bressan01, norton01, yang04, kaviraj07, french18}. Many of these systems show bulge-dominated morphologies, with tidal features or close companions, which suggests a late-stage merger \citep{canalizo00, yang04, goto04, cales11}. Since the SF in these systems is completely quenched, any observed outflows can be attributed solely to the AGN. In addition, due to their narrow stellar age distribution, their starburst age can be used as a clock (see e.g., \citealt{wild10, french18}), where one can study the evolution of outflow properties over hundreds of Myrs. 

Various studies suggest that post starburst E+A galaxies are the evolutionary link between gas rich major mergers (ULRIGs) and quiescent, early-type, galaxies \citep{yang04, yang06, kaviraj07, wild09, cales11, cales13, yesuf14, cales15, french15, wild16, baron17b, baron18, li19}. According to simulations, a gas-rich major merger triggers a powerful starburst, and gas is funneled to the vicinity of the SMBH, triggering an AGN. Soon after, the AGN launches nuclear winds which sweep-up the gas in the galaxy, shutting down the current starburst abruptly, and removing the gas from the host galaxy (e.g., \citealt{springel05, hopkins06}). The more recent simulations suggest that AGN-driven outflows have a limited impact on the ISM of the host galaxy (e.g., \citealt{gabor14, hartwig18}). Instead, it is suggested that a more significant effect is heating of the CGM which prevents further gas accretion, and thus quenches SF in the host galaxy (e.g., \citealt{bower17, pillepich18}). In addition, recent observational evidence challenge the simple picture, by finding large molecular gas reservoirs in such systems (e.g., \citealt{french15, french18}), and finding a significant delay between the onset of the starburst and the peak of AGN activity (e.g, \citealt{wild10, cales15, yesuf14}). For example, \citet{yesuf17} show that post starburst galaxies have a wide range of gas fractions, some are gas rich and some are gas poor. \citet{french18} discover a statistically-significant decline in the molecular gas to stellar mass fraction with the post starburst age. They argued that the implied rapid gas depletion rate of 2--150 $\mathrm{M_{\odot}/yr}$ cannot be due to current SF or supernova feedback, but rather due to AGN feedback (see also \citealt{li19}).

We have recently detected the first evidence of an AGN-driven outflow, traced by ionized gas, in a post starburst E+A galaxy (\citealt{baron17b}; see also \citealt{tremonti07}, \citealt{tripp11}, and \citealt{yesuf17b}). SDSS J132401.63+454620.6 was discovered as an outlier by the anomaly detection algorithm of \citet{baron17}, and our ESI/Keck spectroscopy revealed a post starburst system with powerful ionized outflows, with a mass outflow rate in the range 4--120$\mathrm{M_{\odot}/yr}$. Since then, we have constructed a sample of such galaxies, with fully-quenched starbursts and powerful AGN-driven winds. In a second post starburst E+A galaxy observed with the Keck Cosmic Web Imager (KCWI) on Keck \citep{baron18}, we found that while the stellar continuum is detected to about 3 kpc from the BH, we detect gas outflows to a distance of at least 17 kpc. Our models suggest that the ionized gas outside the galaxy forms a continuous flow, and its mass, roughly $10^{9}\,\mathrm{M_{\odot}}$, exceeds the total gas mass within the galaxy. This suggests that we are witnessing a short-lived phase, in which the AGN is successfully removing most of the gas from its host galaxy.

In this work we present spatially resolved spectroscopy, obtained with MUSE/VLT, of a third E+A galaxy at z=0.090, SDSS J124754.95-033738.6. Our observations reveal galactic-scale ionized and neutral outflows, where we find a remarkable similarity between the spatial extents and kinematics of the two phases. We describe the observations in section \ref{s:observations}, and discuss the general observed properties of the system in section \ref{s:physical_props}. We then describe a single model that accounts for both the neutral and the ionized outflow phases in section \ref{s:models}. We summarize and  conclude in section \ref{s:concs}. Throughout this paper we assume a cosmology with $\Omega_{\mathrm{M}}=0.3$, $\Omega_{\Lambda}=0.7$, and $h=0.7$, thus 1'' corresponds to 1.68 kpc for the system in question. This paper concerns only with the optical properties of the source and further study of the infrared is delayed to future publications.

\section{Observations}\label{s:observations}

\subsection{1D spectroscopy from SDSS}\label{label:s_sdss_1d}

\begin{figure*}
\includegraphics[width=0.95\textwidth]{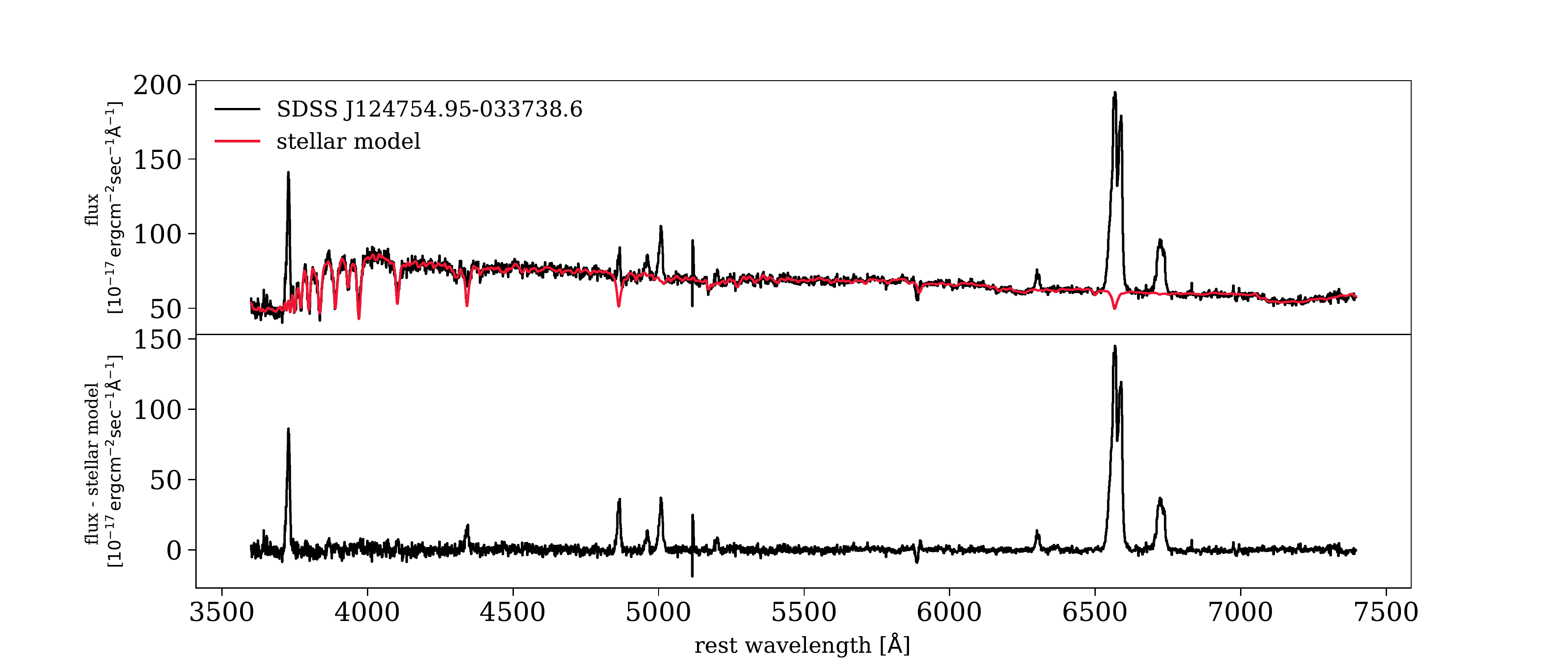}
\caption{Top panel: the SDSS 1D spectrum of SDSS J124754.95-033738.6 (black) and the best-fitting stellar population synthesis model (red) using pPXF (see section \ref{s:stellar_props} for details). Bottom panel: residual spectrum showing the emission and absorption line spectrum.}\label{f:sdss_1d_spectrum_and_fit}
\end{figure*}

SDSS J124754.95-033738.6 was observed as part of the general SDSS survey \citep{york00}. The spectrum was obtained using a 3'' fiber with the SDSS spectrograph, covering the wavelength range 3800\AA--9200\AA, with a resolving power of 1500 at 3800\AA\, and 2500 at 9000\AA. The publicly-available spectrum of the galaxy is combined from three 15m sub-exposures, with the signal-to-noise ratio (SNR) per resolution element of the combined spectrum ranging from 10 to 50. We did not find spectral differences between the 3 sub-exposures. In figure \ref{f:sdss_1d_spectrum_and_fit} we show the 1D SDSS spectrum and its best-fitting stellar model (see section \ref{s:stellar_props} for details about the stellar modeling). 

\subsection{Spatially-resolved spectroscopy with MUSE}\label{label:s_muse_2d}

\begin{figure}
\includegraphics[width=3.5in]{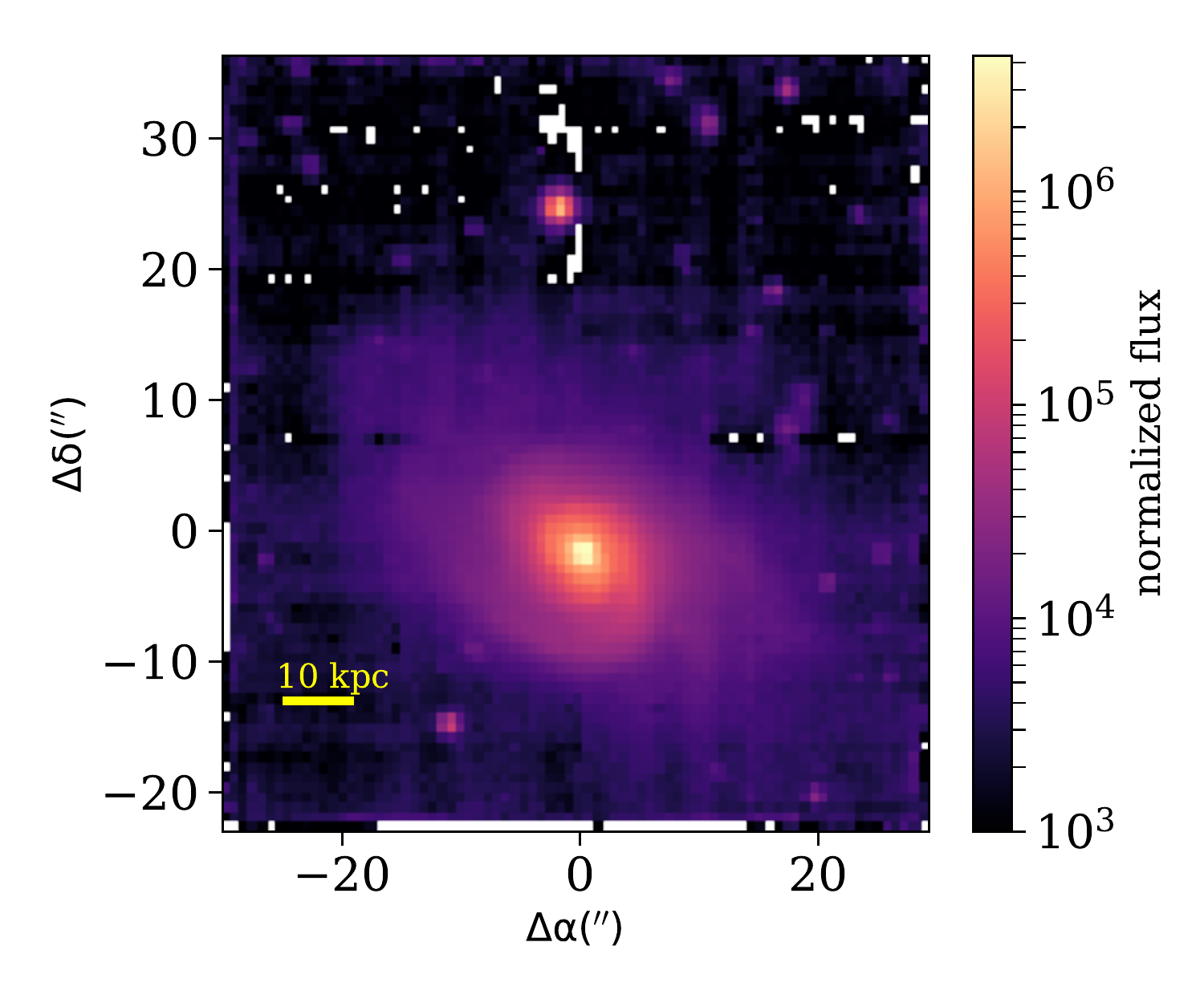}
\caption{\textbf{Integrated stellar continuum emission in the rest-frame wavelength range 5300--5700\AA\, obtained from the MUSE data cube.} The yellow line at the bottom left represents 10 kpc at the redshift of the galaxy ($z = 0.09$). The SNR of the pixels in this image ranges between 100 and 500.}\label{f:muse_continuum_emission}
\end{figure}

MUSE is a second generation integral field spectrograph on the VLT \citep{bacon10}. It consists of 24 integral field units that cover the wavelength range 4650\AA--9300\AA, achieving a spectral resolution of 1750 at 4650\AA\, to 3750 at 9300\AA. In its Wide Field Mode (WFM), MUSE splits the 1$\times$1 arcmin$^{2}$ field of view (FOV) into 24 channels which are further sliced into 48 15''$\times$0.2'' slitlets. SDSS J124754.95-033738.6 was observed as part of our program "Mapping AGN-driven outflows in quiescent post starburst E+A galaxies" (0100.B-0517(A); P.I. R. Davies). The observations were performed over two observing blocks (OBs) on March 9 and April 7 2018, in a seeing-limited WFM with a pixel scale of 0.2''. The total exposure time of the combined MUSE cube is 2.0 hours, with an average spatial resolution of FWHM=0.71''. We downloaded the data from the ESO phase 3 online interface, which provides fully reduced, calibrated, and combined MUSE data for all targets with multiple OBs. Given the spaxel scale of the WFM and the spatial resolution of the combined cube, we obtain our final data cube by summing $3 \times 3$ spaxel regions in the original cube. At the redshift of the system ($z = 0.09$), each spaxel of 0.6'' in the final cube represents 1 kpc in the galaxy. The resulting signal to noise ratio (SNR) of spaxels where the source is detected ranges between 5 to 25 per spatial and wavelength resolution element. In figure \ref{f:muse_continuum_emission} we show the integrated stellar continuum emission in the rest-frame wavelength range 5300--5700\AA\, using the final MUSE data cube. The image shows a spiral galaxy with two clear spiral arms, extending to distances of more than 20 kpc from the center. 

\subsection{X-ray observations with XMM-Newton}\label{label:s_xray}

SDSS J124754.95-033738.6 was observed by \emph{XMM-Newton} (\citealt{jansen01}) on August 14 2018, as part of our program "X-ray properties of quiescent post starburst galaxies with AGN-driven winds" (AO17 82008; P.I. H. Netzer), in which we observed six post starburst E+A galaxies with massive AGN-driven winds. The \emph{XMM-Newton} instrument modes were full-frame, with a thin filter for the three EPIC cameras, and the optical/UV filters U, UVW1, and UVM2 for the OM\footnote{See additional details at: \url{https://xmm-tools.cosmos.esa.int/external/xmm_user_support/documentation/uhb/omfilters.html}}. The total observation duration for EPIC was 12 ks, and the exposure time for OM was 6 ks. The raw data files were processed, using standard procedures, with version 15.0 of the \emph{XMM-Newton} Scientific Analysis Software (SAS2) package, with the 2017 release of the Current Calibration File. Source events were extracted from a circular region of 20 arcsec in radius, and background events from a region of 1 arcmin in radius, which was away from the source and clean from other sources. 

\section{Physical properties}\label{s:physical_props}

In this section we present the physical properties of SDSS J124754.95-033738.6. We study the stellar properties of the system in section \ref{s:stellar_props}, the AGN properties in section \ref{s:agn_props}, and the gas properties (ionized and neutral) in section \ref{s:gas_props}.

\begin{figure*}
\includegraphics[width=0.95\textwidth]{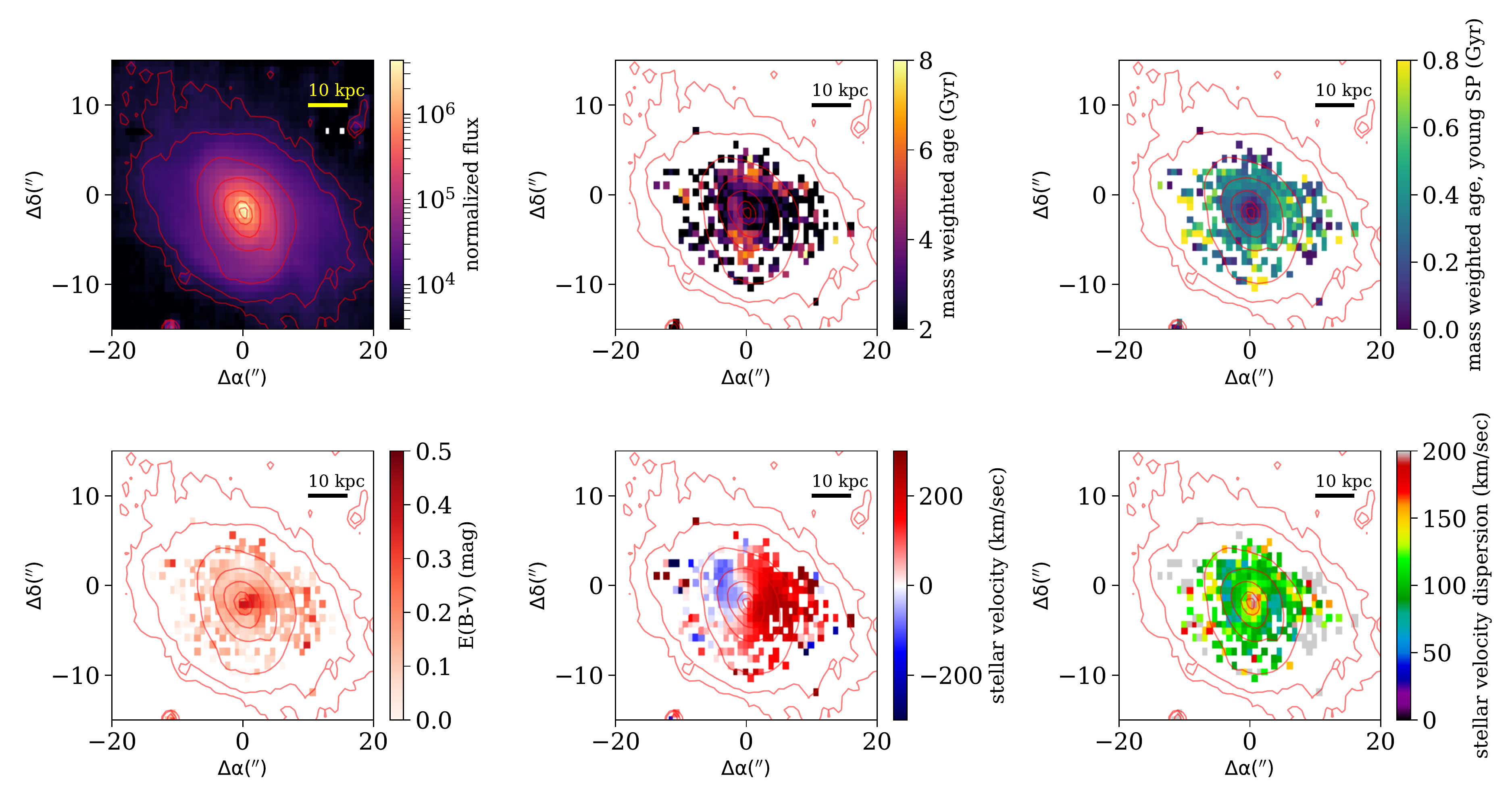}
\caption{\textbf{Spatially-resolved stellar properties, obtained from {\sc pPXF} fitting of individual spaxels.} The top panels show (from left to right): integrated stellar continuum in the rest-frame wavelength range 5300--5700\AA, mass-weighted age of the best-fitting stellar model, and mass-weighted age of stars younger than 1 Gyr. The bottom panels show (from left to right): dust reddening towards the stars, stellar velocity with respect to systemic velocity, and stellar velocity dispersion. The red contours represent different integrated stellar continuum values in the rest-frame wavelength range 5300--5700\AA, ranging from $10^{7}$ to $10^{4}$ in logarithmic steps of 0.5 dex, with physical units of $10^{-20}\, \mathrm{erg/(sec\, cm^{2} \AA)}$. }\label{f:2d_stellar_properties}
\end{figure*}

\subsection{Stellar properties}\label{s:stellar_props}

In this section we use the 1D and spatially-resolved spectra to study the stellar properties of the source, to constrain the star formation history, and to measure the stellar kinematics and the reddening towards the stars.

The 1D SDSS spectrum of SDSS J124754.95-033738.6 (figure \ref{f:sdss_1d_spectrum_and_fit}) is dominated by A-type stars and shows strong Balmer absorption lines, which is a clear signature of post starburst E+A galaxies. We used the 1D SDSS spectrum to measure the equivalent width (EW) of the H$\delta$ absorption line and found EW(H$\delta$)=6.7\AA, which is above the typical 5\AA\, threshold used to select post starburst E+A galaxies (e.g., \citealt{goto07, alatalo16a}). We then fitted a stellar population synthesis model using the {\sc python} implementation of Penalized Pixel-Fitting stellar kinematics extraction code ({\sc pPXF}; \citealt{cappellari12}). {\sc pPXF} is a public code for extraction of the stellar kinematics and stellar population from absorption line spectra of galaxies \citep{cappellari04}. Its output includes the best-fitting stellar model, the relative contribution of stars with different ages, the stellar velocity dispersion, and the dust reddening towards the stars (assuming a \citealt{calzetti00} extinction law). The code uses the MILES library, which contains single stellar population synthesis models that cover the entire optical wavelength range with a FWHM resolution of 2.3\AA\,\citep{vazdekis10}. We used models produced with the Padova 2000 stellar isochrones assuming a Chabrier initial mass function (IMF; \citealt{chabrier03}). The stellar ages range from 0.03 to 14 Gyr, thus allowing the analysis of systems with different star formation histories. The best-fitting stellar model is marked with red in figure \ref{f:sdss_1d_spectrum_and_fit}. Its stellar age distribution consists of two star formation episodes, with a recent short episode that started 70 Myrs ago and was quenched 30 Myrs ago, and an older long episode that started 14 Gyrs ago and ended 6 Gyrs ago. The older episode is poorly constrained and we find similarly reasonable fits when forcing the code to use templates with different ages within the range 3--10 Gyrs. According to the best-fitting model, the stellar velocity dispersion is 185 km/sec and the dust reddening towards the stars is $E_{\mathrm{B-V}} = 0.188$ mag. The total stellar mass is $\mathrm{M_{*}} = 10^{10.8}\,M_{\odot}$, which is consistent with other estimates (e.g., \citealt{chen12}). About $\sim$2\% of this mass was formed during the recent burst. 

\begin{figure}
\includegraphics[width=3.5in]{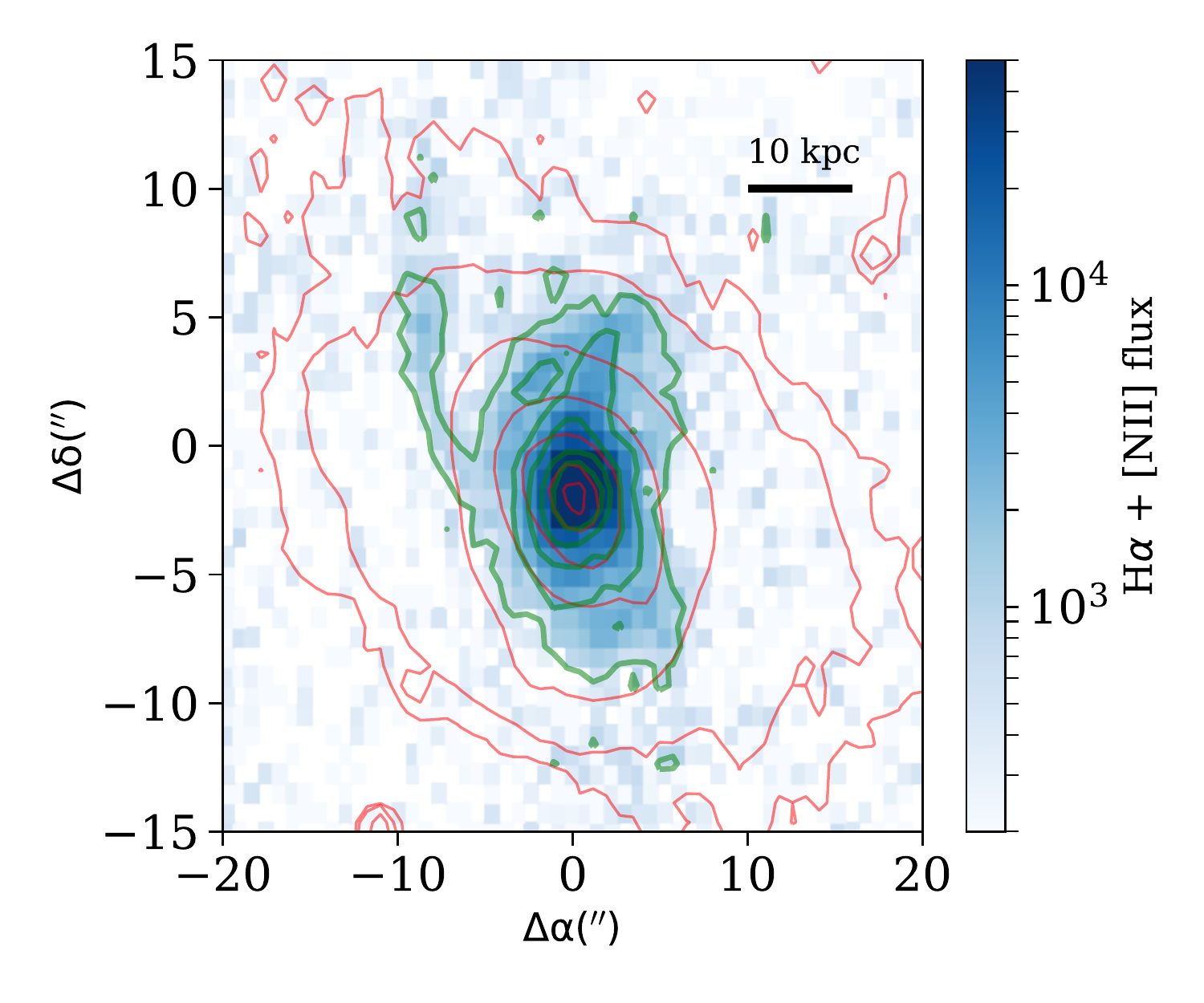}
\caption{\textbf{Ionized gas and stellar emission distribution.} The colors represent the integrated line emission in the rest-frame wavelength range 6530--6600\AA, which includes the H$\alpha$ and [NII]~$\lambda \lambda$ 6548,6584\AA\, emission lines. The red contours represent steps of 0.5 dex in stellar continuum emission (see figure \ref{f:2d_stellar_properties}). The green contours represent different integrated H$\alpha$+[NII] flux values, ranging from $10^{5}$ to $10^{3}$ in logarithmic steps of 0.5 dex, with physical units of $10^{-20}\, \mathrm{erg/(sec\, cm^{2} \AA)}$. Clearly the ionized gas is disturbed and does not follow the stellar emission.}\label{f:ionized_gas_emission}
\end{figure}

The integrated stellar continuum emission (using the MUSE cube; figure \ref{f:muse_continuum_emission}) indicates that the system is a spiral galaxy, with at least two spiral arms which extend to distances of more than 20 kpc from the center of the galaxy. The ordered and symmetric stellar distribution in SDSS J124754.95-033738.6 argues against the system being a product of a major merger, since such systems usually show disturbed and asymmetric stellar distributions. One might argue that this is at odds with our spectral classification of this galaxy as an E+A galaxy, since many E+A galaxies show tidal features or close companions, which are suggestive of a late-stage merger (e.g., \citealt{canalizo00, yang04, goto04, cales11}). However, \citet{cales11}, who studied a sample of 29 post starburst quasars using images from the Hubble Space Telescope, found an equal number of spiral (13/29) and early-type (13/29) host galaxies. Thus, post starburst E+A spectral signatures can also be present in spiral galaxies that show no sign of a major morphological disturbance, where the starburst might have been triggered by a minor merger (see section \ref{s:gas_props} and figure \ref{f:ionized_gas_emission} for an additional indication for a minor merger).  

To obtain the spatially-resolved stellar properties of the system, we fitted stellar population synthesis models to the individual spaxels in the MUSE data cube. In figure \ref{f:2d_stellar_properties} we present the results of our {\sc pPXF} fits. The top panel shows the integrated stellar continuum in the rest-frame wavelength range 5300--5700\AA, the mass-weighted age of the best-fitting stellar model, and the mass-weighted age of stars younger than 1 Gyr. Similar to the 1D SDSS spectrum, all spaxels show contributions from two separate star formation episodes, one considerably younger than 1 Gyr and one which is older than 1 Gyr. The top middle panel in figure \ref{f:2d_stellar_properties} represents the distribution of all stellar ages, and is thus more sensitive to the older stellar population. The top right panel in the figure represents the distribution of stars younger than 1 Gyr, and is therefore sensitive only to the recent star formation episode. The bottom panels of figure \ref{f:2d_stellar_properties} show the dust reddening towards the stars, the stellar velocity with respect to the systemic velocity, and the stellar velocity dispersion. We also use red contours to mark steps of 0.5 dex in stellar continuum emission in all the panels. To facilitate the comparison between the different diagrams, we use the same contours in most of the figures shown the paper.

The stellar continuum emission shows no evidence for ongoing SF throughout the FOV. However, in section \ref{s:gas_props} we perform emission line decomposition and show that the emission line ratios in the central spaxels are consistent with composite spectra, suggesting that the gas is ionized by a combination of AGN and SF radiation. To estimate the SFR and its relative contribution to the ionization of the different lines, we use the prescription by \citet{wild10} and the narrow kinematic components of the emission lines: [OIII], H$\beta$, [NII], and H$\alpha$ (see additional details in section \ref{s:gas_props}). We find that the relative contribution of SF to the H$\alpha$ luminosity ranges from 0.6 to 1 in the few central spaxels. Assuming Kroupa IMF and using the galaxy-integrated dust-corrected $\mathrm{L_{SF}(H\alpha)}$, we find SFR of $2.2\,\mathrm{M_{\odot}/yr}$. Although non-negligible, this SFR places this system 0.4 dex (roughly 1$\sigma$) below the SF main sequence at z=0.1 \citep{whitaker12}. Without additional far-infrared data, we cannot say anything about the mid-infrared properties of the outflow \citep{baron19a} and the possibility of heavily obscured SF regions.

\subsection{AGN type and bolometric luminosity}\label{s:agn_props}

The EPIC PN spectrum of SDSS J124754.95-033738.6 is consistent with a typical low-SNR unobscured type I AGN. To fit the X-ray spectrum, we used the {\sc XSPEC} task \citep{arnaud96} version 12.8.2. To obtain a similar SNR for the different spectral bins, we grouped the PI channels of the source-spectra, using the task \emph{grppha} of the {\sc FTOOLS} \citep{blackburn95} version 6.10. We fitted the spectrum with an absorbed power-law with photon index $\Gamma$ and an obscuring column of $\mathrm{N_{H}}$. The best-fitting power-law has $\Gamma=1.801 \pm 0.094$, and a normalization of $1.66 \times 10^{-05}  \pm 2.0 \times 10^{-06}\,\mathrm{photons\,cm^{-2}\,sec^{-1}\,keV^{-1}}$. The best-fitting column density is $\mathrm{N_{H} = 5.6 \times 10^{20}  \pm 3.6 \times 10^{20}\, cm^{-2}}$, consistent with galactic obscuration. Using the best-fitting model, we get $\mathrm{f(2-10\,keV) = 3 \times 10^{-13}\, erg/cm^{2}/sec}$, corresponding to $\mathrm{L(2-10\,keV) = 6 \times 10^{42}\, erg/sec}$.

We estimate the bolometric luminosity of the AGN. The accretion disk (AD) calculations by \citet{netzer19} suggest a 2--10 keV bolometric correction factor in the range 10--30 with a mean value of $\mathrm{k_{bol}(2-10\,keV)} \sim 11$. However, the resulting bolometric luminosity is an order of magnitude lower than the AGN luminosity derived from the narrow emission lines [OIII], H$\beta$, and [OI] (see \citealt{netzer09}, and additional details in section \ref{s:origin_of_narrow_lines}). As a compromise, we choose $\mathrm{k_{bol}(2-10\,keV)} \sim 25$, which is within the range suggested by \citet{netzer19} and is also consistent with possible variations of the X-ray continuum given that the the emission line luminosity represents the long-term average of the optical-UV and X-ray continuum. This gives $\mathrm{L_{AGN} = 1.5 \times 10^{44}\, erg/sec}$, which is probably correct to a factor of 5. 

The optical spectrum of SDSS J124754.95-033738.6 is dominated by stellar continuum with no evidence for an AGN continuum emission in these wavelengths. We considered the possibility of a Compton-thick type II AGN, where the observed spectrum is dominated by reflection. However, the reddening-corrected narrow [OIII] luminosity is $\mathrm{L([OIII]) = 3 \times 10^{41}\,\mathrm{erg/sec}}$, resulting in a luminosity ratio $\mathrm{L_{2-10\,keV}/L_{[OIII]} \sim 20}$. This contradicts the results in figure 1 from \citet{bassani99}, which shows that Compton-thick type II AGN have $\mathrm{L_{2-10\,keV}/L_{[OIII]} < 1}$. We thus conclude that this source is consistent with a typical unobscured type I AGN. 

The reddening we derive towards the stars and gas in the central spaxel is $\mathrm{E}_{B-V}$ of 0.5--0.8 mag (see details in section \ref{s:gas_props}), which, for a standard gas-to-dust ratio, would give $\mathrm{N_{H} > 3 \times 10^{21}\, cm^{-2}}$. Furthermore, \citet{maiolino01} suggested that the gas-to-dust ratio in some AGN is larger than the standard value, leading to an even larger obscuring column (see also \citealt{burtscher16} who suggested that this is due to additional obscuration by BLR clouds, and references therein). These obscuring columns are inconsistent with the one derived from the X-ray fitting. The different columns might be related to the different components originating from different lines of sight: since the diameter of the central spaxel is $\sim$1 kpc, it is possible that the vicinity of the BH is unobscured, while the stars and gas in the central 1 kpc are.

Since the source is classified as an unobscured type I AGN, we expect to detect blue AD continuum and broad Balmer emission lines. We calculated the expected contributions from these components using the derived X-ray luminosity and the expression given in \citet{netzer19}. We find $\mathrm{k_{bol}(5100\,\AA)}=28$. Therefore, the AD continuum emission is expected to be $\mathrm{\nu L_{\nu} (5100\AA) = 5.3 \times 10^{42}\, erg/sec}$. The dust-corrected stellar continuum luminosity at the central spaxel at $\lambda=5100$\AA\, is larger than this value by an order of magnitude. This suggests that the AD cannot be detected even in the bluest wavelengths of the optical spectrum. 

Next, we examine whether we expect to detect the wings of a broad H$\alpha$ line that originates in the BLR. We assume that the broad H$\alpha$ is described by a Gaussian profile with a typical FWHM=4\,500 km/sec\footnote{The conclusions remain unchanged for FWHM=3\,000 km/sec.}. The emission line is obscured by dust with dust reddening of $\mathrm{A_{V}}=0.3$ mag, consistent with the obscuring column derived from the X-ray fitting. We do not detect any broad wings that originate from a broad H$\alpha$ line in the central source. This places an upper limit on the broad H$\alpha$ luminosity: $3 \times 10^{41}$ erg/sec. We compare this limit to the expected line luminosities using scaling relations from the literature. These relations tie the optical continuum emission or the X-ray luminosity to the broad H$\alpha$ luminosity. Using the relations by \citet{greene05}, \citet{panessa06}, and \citet{stern12}, we expect the broad H$\alpha$ luminosity to be: $5 \times 10^{40}$ erg/sec, $1 \times 10^{41}$ erg/sec, or $5 \times 10^{41}$ erg/sec respectively. The upper limit we deduce is consistent with most of the relations cited above, suggesting that the broad lines are not detected due to the significant stellar continuum emission. 

To summarize, we suggest that the system hosts an unobscured type I AGN. We do not detect blue AD continuum and broad Balmer emission lines in optical wavelengths due to the significant stellar radiation in the central spaxel of our source.

\subsection{Gas properties}\label{s:gas_props}

In this section we study the properties of the ionized and neutral gas in the system. In section \ref{s:model_inedpendent_props} we present general observed properties. In particular, we present evidence for disturbed ionized gas morphology that is suggestive of a minor-merger, and evidence of spatially-resolved NaID emission and absorption. We further discuss the kinematic connection between the neutral and the ionized gas phases. In section \ref{s:emis_decomp} we present our method of emission and absorption line decomposition, and suggest a novel treatment of the NaID emission and absorption complex. In section \ref{s:derived_gas_props} we present the derived ionized and neutral gas properties. For the ionized gas, we study its spatial distribution and kinematics, the main source of ionization, the gas ionization state, reddening, and electron density. For the neutral gas, we study its spatial extent and kinematics, its emission luminosity and absorption equivalent width (EW), and its optical depth and covering factor. Finally, in section \ref{s:connection_between_gas_phases} we discuss the connection between the neutral and ionized gas phases in the system.

\subsubsection{General observed properties}\label{s:model_inedpendent_props}

We detect various emission and absorption lines, tracing both ionized and neutral gas phases, throughout the entire FOV. To examine the emission lines, we subtract the best-fitting stellar model from each spaxel, resulting in an emission line cube. In figure \ref{f:ionized_gas_emission} we show the integrated flux in the rest-frame wavelength 6530--6600\AA\, of the emission line cube, a range that includes the H$\alpha$ and [NII]~$\lambda \lambda$ 6548,6584\AA\, emission lines. Contrary to the ordered and symmetric stellar distribution we found in section \ref{s:stellar_props}, the line emitting gas appears to be disturbed and asymmetric, with a clear tidal tail extending to the north-east (top left) direction. In figure \ref{f:oiii_and_hbeta_chanel_maps} in the appendix we show the integrated H$\beta$ and [OIII]~$\lambda \lambda$ 4959,5007\AA\, flux for different velocity channels, where we detect both narrow and broad emission lines throughout the FOV. The disturbed gas morphology observed in both of these diagrams reinforces the suggestion that SDSS J124754.95-033738.6 is an E+A galaxy that has experienced a minor merger in its recent past. The post starburst E+A spectral signatures might be the result of a star formation burst that was triggered during the accretion of a companion galaxy (see additional example by \citealt{cheung16}). 

\begin{figure*}
\includegraphics[width=0.95\textwidth]{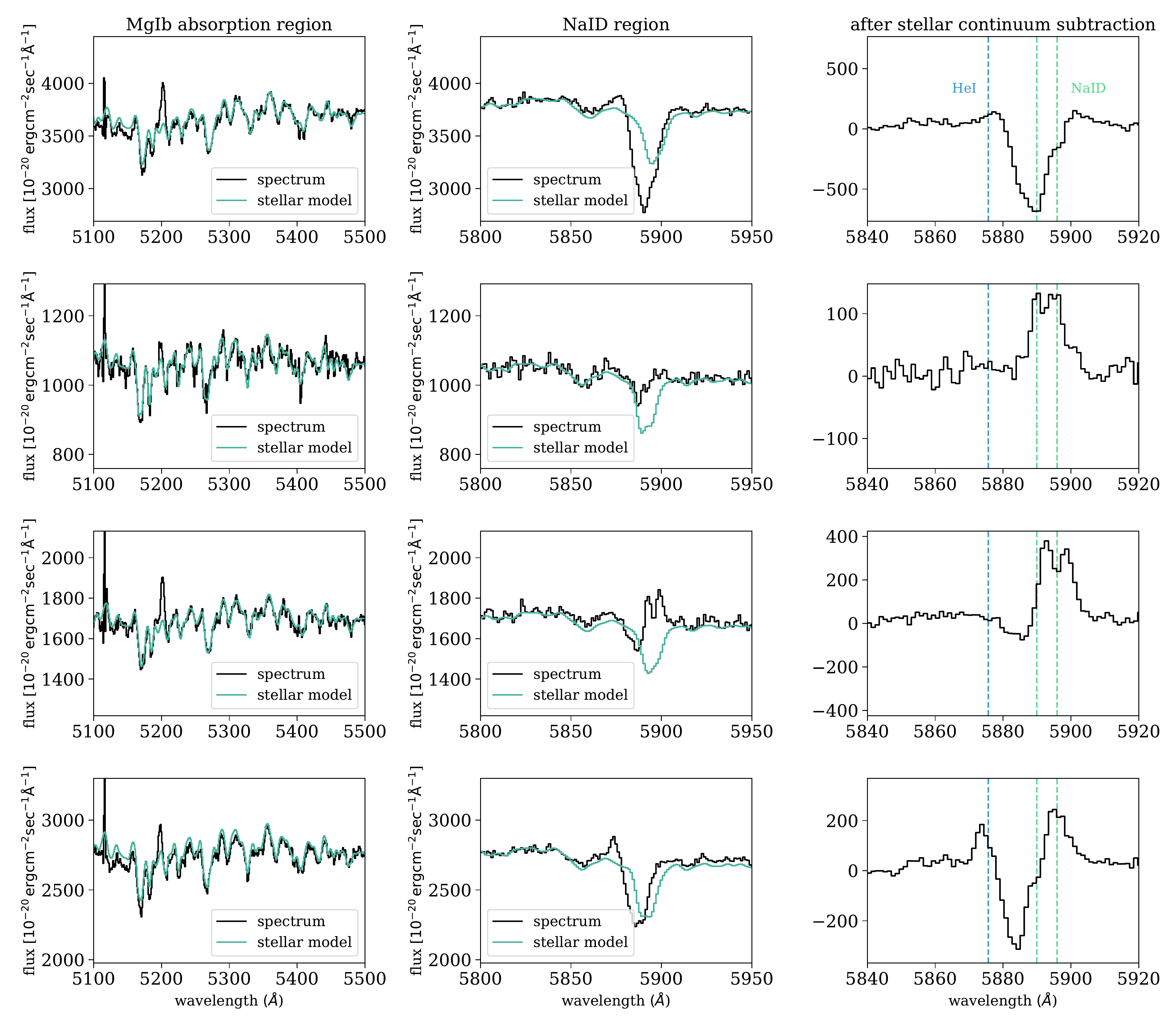}
\caption{\textbf{Evidence of NaID emission.} Four representative spaxels (out of a total of 80 spaxels) in which we detect NaID emission. Each row represents a different spaxel. The left panels show the observed spectra, centered around the stellar MgIb line, and their best-fitting stellar population synthesis models. The middle panels show the observed spectra in the NaID region with their corresponding best-fitting stellar models. The right panels show the spectra after subtracting the best-fitting stellar models, centered around the NaID region. We mark the systemic wavelength of the HeI emission and NaID absorption in the right panels. }\label{f:naid_emission_example_1}
\end{figure*}

\begin{figure*}
\includegraphics[width=1\textwidth]{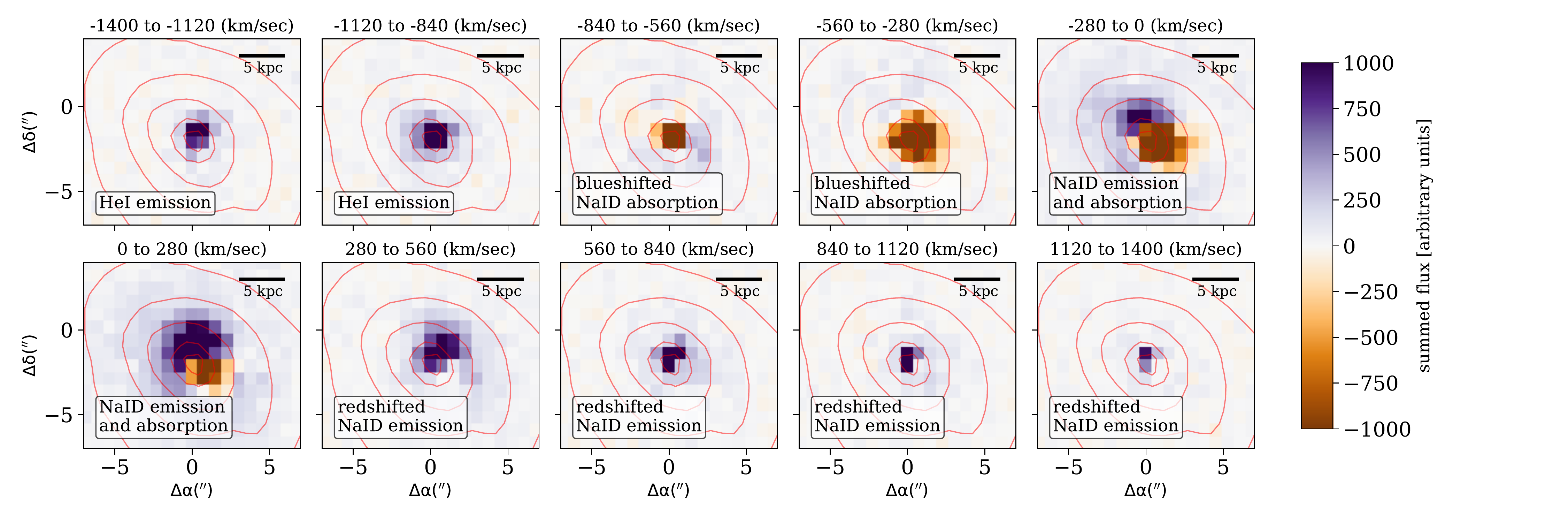}
\caption{\textbf{Spatial distribution of the NaID emission and absorption.} The different panels show different velocity channels, ranging from -1400 km/sec to +1400 km/sec. Since we integrate over the residual flux, emission lines are represented by positive values (purple colors) and absorption lines by negative values (orange colors). The NaID-emitting gas reaches very large velocities of $-$800 km/sec and +1400 km/sec.
}\label{f:naid_emission_example_2}
\end{figure*}

We detect resonant NaID absorption and \textbf{emission} throughout a large fraction of the FOV. Blueshifted NaID absorption that traces cool neutral outflows has been detected and analyzed in numerous systems with and without an AGN (e.g., \citealt{rupke05c, martin06, shih10, rupke13, cazzoli16, rupke17} and references therein). While such features have  been predicted to exist \citep{prochaska11}, resonant NaID emission has only been detected in a handful of sources so far (\citealt{phillips93, rupke15, perna19}; see also \citealt{rubin10, rubin11} for detection of resonant MgII emission). SDSS J124754.95-033738.6 is the third reported system in which this emission is spatially-resolved. 

In figure \ref{f:naid_emission_example_1} we show four representative spaxels (out of a total of 80) in which we detect NaID emission. The left panels show the observed spectra, centered around the stellar MgIb line, and their best-fitting stellar population synthesis models. The stellar population synthesis model fits the stellar continuum and the stellar absorption lines very well. The middle panels show the observed spectra in the NaID region, where there is a significant disagreement between the spectrum and the best-fitting stellar model. This disagreement remains unchanged for different choices of stellar model parameters, like stellar ages, stellar initial mass function, and isochrones. Since the strengths of stellar MgIb and stellar NaID are proportional to each other (\citealt{heckman00, rupke02}), the agreement in the MgIb region and the disagreement in the NaID region suggest an additional NaID absorption and emission component. In the right panels of figure \ref{f:naid_emission_example_1} we show the spectra after subtracting the best-fitting stellar models, and mark the systemic velocity of the HeI emission line and NaID absorption lines. Clearly, there is a redshifted NaID emission that cannot be mistaken for HeI emission, which is mixed with broad and blueshifted NaID absorption. In section \ref{s:emis_decomp} below, we perform a global emission and absorption line decomposition into all of these components. 

It is informative to examine the spatial distribution of the NaID emission and absorption with respect to the stellar continuum, in a way which is independent of our adopted line decomposition technique. To do so, we subtract the best-fitting stellar models from all the spaxels in our cube, resulting in an emission line cube. We then integrate the residual flux in wavelength bins that correspond to different velocity channels with respect to the systemic NaID absorption. In figure \ref{f:naid_emission_example_2} we show the result of this process, for velocity channels in the range -1400 km/sec to +1400 km/sec. Since we integrate over the residual flux, emission lines are represented by positive integrated flux values and absorption lines are represented by negative integrated flux values\footnote{We note that such a representation is not physically meaningful since an absorption line is a multiplicative, rather than additive, effect, and the negative flux values have no clear physical meaning. A physically-meaningful representation requires emission and absorption line decomposition which requires model assumptions and is subjected to various fitting degeneracies.}. Inspecting the panels from left to right and from top to bottom reveals: HeI emission close to systemic velocity (panels 1-2), blueshifted NaID absorption (panels 3-4), a combination of NaID emission and absorption close to systemic velocity (panels 5-6), and redshifted NaID emission (panels 7-10). Finally, one can see that the NaID absorption and emission occupy roughly the same spatial regions in the galaxy, suggesting that most of the spaxels in the cube consist of a combination of NaID absorption and emission, which we further discuss below. 

Most of the spaxels show a combination of narrow and broad HeI emission (similarly to the other ionized emission lines), narrow and broad NaID emission, and broad blueshifted NaID absorption (see e.g. figure \ref{f:emis_and_abs_fitting_example_neutral_only_with_explanations}). We also find a surprising similarity between the kinematics of the different NaID components and the kinematics of different ionized emission components. We find similar velocities and velocity dispersions for the narrow NaID emission lines and the narrow ionized lines. The broad blueshifted NaID absorption shows a similar kinematic profile to the blueshifted wing of the broad ionized emission lines, and the broad redshifted NaID emission shows a similar profile to the redshifted wing of the broad ionized emission lines. These similarities suggest that the NaID and the ionized lines originate from the \emph{same} outflowing clouds, which are distributed in a double-cone or an outflowing shell-like geometry. In such a case, the approaching side of the outflow will contribute blueshifted ionized emission and blueshifted NaID absorption, and the receding part of the outflow, if not too extincted by dust, will contribute redshifted ionized emission (which we observe) and redshifted NaID emission. 

Here, and in other cases where NaID is observed in emission, the redshifted NaID emission is the result of absorption of continuum photons by NaI atoms in the receding side of the outflow and isotropic reemission. Such photons are redshifted with respect to systemic velocity and are not absorbed by NaI atoms in the approaching side of the outflow (e.g., \citealt{prochaska11}). This results in a classical P-Cygni profile. Therefore, the observed NaID emission and absorption profile in our source suggests a neutral gas outflow, where we observe both the approaching gas through blueshifted absorption and the receding gas through redshifted emission. In this configuration, the gas that produces the blueshifted NaID absorption is \emph{in front} of the gas that produces the redshifted emission. Since the neutral and ionized lines originate from the same outflowing clouds, this also suggests that the gas that produces the NaID absorption is \emph{in front} of the ionized gas that produces the redshifted emission. We use these observations in section \ref{s:emis_decomp} below to construct a model for the emission and absorption in this system.

\subsubsection{Emission and absorption line decomposition}\label{s:emis_decomp}

We detect ionized line emission in 259 spaxels throughout the FOV. Out of these, we detect additional broad ionized kinematic components in 73 spaxels, and NaID emission and absorption in 80 spaxels. The MUSE rest-frame wavelength range allows us to observe the following ionized emission lines: $\mathrm{H\beta}$~$\lambda$ 4861\AA, $\mathrm{HeI}$~$\lambda$ 5876\AA, $\mathrm{H\alpha}$~$\lambda$ 6563\AA, [OIII]~$\lambda \lambda$ 4959,5007\AA, [OI]~$\lambda \lambda$ 6300,6363\AA, [NII]~$\lambda \lambda$ 6548,6584\AA, and [SII]~$\lambda \lambda$ 6717,6731\AA\, (hereafter $\mathrm{H\beta}$, $\mathrm{HeI}$, $\mathrm{H\alpha}$, $\mathrm{[OIII]}$, $\mathrm{[OI]}$, $\mathrm{[NII]}$, and $\mathrm{[SII]}$).

We start our emission line decomposition with the H$\alpha$ and [NII] emission lines, since these are the strongest emission lines we observe in our emission line cube\footnote{Many studies start the emission line decomposition from the [OIII] line, since it is one of the strongest optical emission lines and since it is not blended. In our source, the [OIII] emission line is much weaker than the H$\alpha$ line due to significant dust extinction, and thus its SNR is smaller than that of the H$\alpha$ line. }. We model each of the emission lines using one or two Gaussians, where the first represents the narrow kinematic component and the second the broader kinematic component. We tie the central wavelengths and widths of the narrow Gaussians to have the same velocity and velocity dispersion, and do the same for the broader Gaussians. We force the [NII] intensity ratio to its theoretical value. The broad kinematic component is kept only if its flux in H$\alpha$ and [NII] is detected to more than $3\sigma$. Otherwise, we perform an additional fit with a single narrow kinematic component. 

Once we obtain the best-fitting model for the H$\alpha$+[NII] complex, we use the best-fitting parameters to constrain the fit of the other ionized emission lines. We fit narrow and broad (if exists in the H$\alpha$+[NII] fit) kinematic components to the H$\beta$, [OI], [OIII], and [SII] lines, where we force their central wavelengths and widths to show the same velocities and velocity dispersions to those we found for the H$\alpha$ and [NII] lines. We further force the [OIII] intensity ratio to its theoretical value, and limit the [SII] intensity ratio, $\mathrm{[SII]\lambda 6717\AA / [SII]\lambda 6731\AA}$, to be in the range 0.44--1.44. 

We show examples of the best-fitting models in figure \ref{f:emis_and_abs_fitting_example_ionized_only} in the appendix, where the total model is marked with a red line, and the narrow and broad kinematic components are marked with green and blue lines respectively. One can see that the two Gaussians model provides an adequate representation of the ionized emission lines. Out of the 259 spaxels we fitted, 186 require only one kinematic component and 73 require two kinematic components. One can see in figure \ref{f:emis_and_abs_fitting_example_ionized_only} that the broad ionized emission lines often show both redshifted and blueshifted wings (with respect to the narrow kinematic component), suggesting that we detect both the approaching and the receding sides of the outflow. One could, in principle, fit these spectra with three kinematic components, one that represents the narrow core of the line, and two that represent the blueshifted and redshifted wings of the line, which are associated with the two sides of the outflow. We attempted to perform such fits and found that the model is too complex, with the different components being somewhat degenerate with each other. In particular, the best-fitting models of the three components 
showed a large variation in emission line ratios and widths between neighboring spaxels, which we believe to not be physical. In the case of the two kinematic components (narrow and broad), we found a continuous change of different line properties (such as line ratios and line widths) across spaxels, without imposing such constrains explicitly. We therefore chose the two component model over the three component one.

Next, we model the HeI+NaID complex. The observed spectrum in this wavelength range includes a contribution from stellar continuum (which contributes to the NaID absorption), narrow and broad HeI emission, narrow and broad redshifted NaID emission, and broad blueshifted NaID absorption. The full model can be expressed as:
\begin{equation}\label{eq:1}
	\begin{split}
	& \mathrm{f_{total}(\lambda) = \Big[ f_{stars}(\lambda) + f_{HeI}(\lambda) + f_{narrow\,NaID\,emis}(\lambda) + } \\
	& \mathrm{+ f_{broad\,NaID\,emis}(\lambda) \Big] \times f_{broad\,NaID\,abs}(\lambda)},
	\end{split}
\end{equation}
where $\mathrm{f_{stars}(\lambda)}$ is the stellar continuum, $\mathrm{f_{HeI}(\lambda)}$ is the narrow and broad HeI emission, $\mathrm{f_{narrow\,NaID\,emis}(\lambda)}$ is the narrow NaID emission, $\mathrm{f_{broad\,NaID\,emis}(\lambda)}$ is the broad NaID emission, and $\mathrm{f_{broad\,NaID\,abs}(\lambda)}$ is the broad NaID absorption. In our adopted model, the NaID absorption affects all the emission components in the system. This is justified by our observation that the gas that produces the blueshifted NaID absorption is in front of the gas that produces the redshifted NaID and ionized emission lines (see section \ref{s:model_inedpendent_props}). Our proposed model is too complex and its parameters cannot be constrained by our observations. In particular, we expect various parameters to be degenerate with each other (e.g., the strength of the HeI emission is degenerate with the strength of the blueshifted NaID absorption). We therefore made several simplifying assumptions that resulted in a simpler model with fewer free parameters. 

We use the best-fitting stellar population synthesis model (see section \ref{s:stellar_props}) as the stellar model, $\mathrm{f_{stars}(\lambda)}$. We model the narrow+broad HeI emission using the best-fitting parameters from the H$\alpha$ line fitting: the velocity and the velocity dispersion of the HeI line are taken to be exactly equal to those of the H$\alpha$ line, which is a very reasonable assumption for these two recombination lines. For case-B recombination, the HeI to H$\alpha$ intensity ratio depends only slightly on the electron temperature, and is 0.033 for $T_{e}=10^{4}$ K \citep{osterbrock06, draine11}. Therefore, the narrow+broad HeI model, $\mathrm{f_{HeI}(\lambda)}$, is determined completely from the best-fitting narrow+broad H$\alpha$ line. Thus, $\mathrm{f_{stars}(\lambda)}$ and $\mathrm{f_{HeI}(\lambda)}$ are determined by other observations and have no free parameters.

\begin{figure*}
\includegraphics[width=0.9\textwidth]{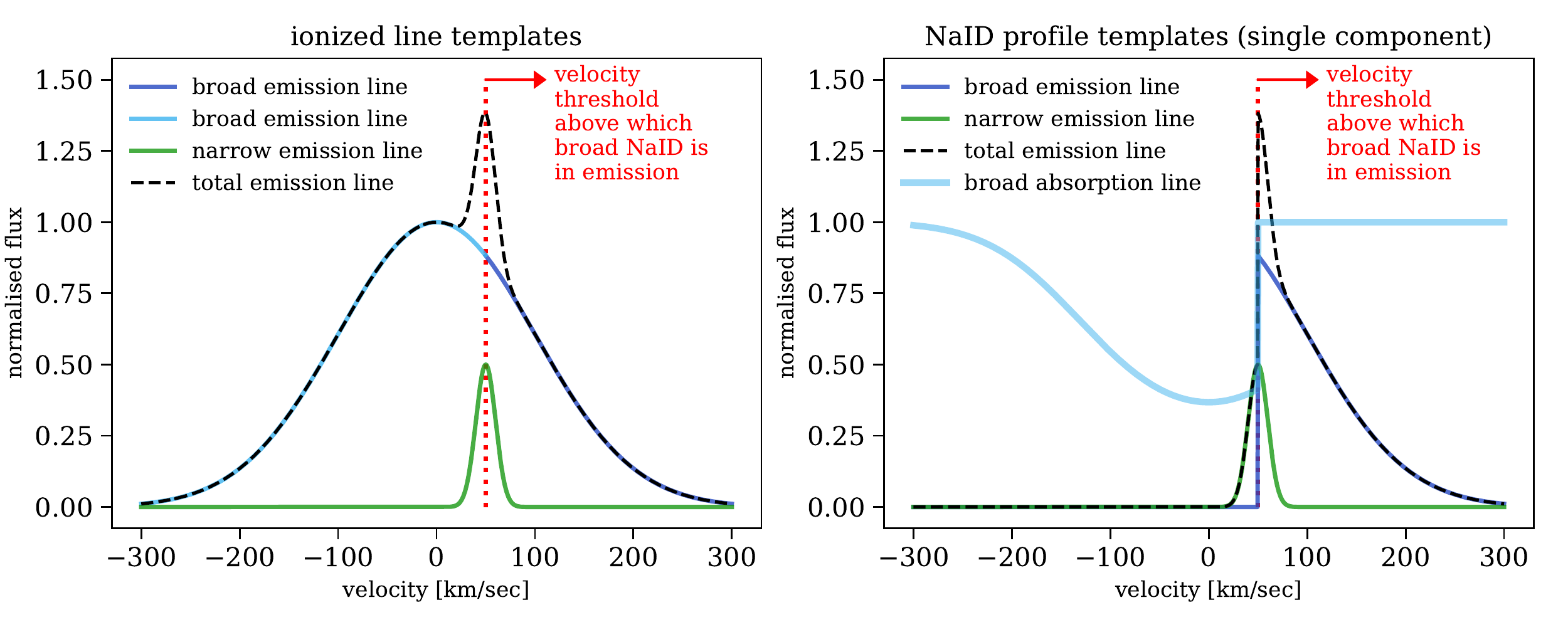}
\caption{\textbf{Illustration of the process used to construct the NaID absorption and emission templates.} The left panel shows an example of a best-fitting emission line profile of one of the ionized lines (black dashed line), which is a combination of a narrow (green line) and a broad (cyan and blue line) kinematic components. The broad emission component is split into its redshifted (blue line) and blueshifted (cyan line) parts with respect to the velocity threshold used to model the NaID emission and absorption. These two represent the receding and approaching sides of the ionized outflow respectively. The right panel shows the resulting templates for the NaID emission and absorption, where narrow NaID emission is marked with green, broad NaID emission is marked with blue, and broad NaID absorption is marked with cyan. The red dotted line in both of the panels represents the velocity threshold that is used to split the ionized emission profile into a template P-Cygni-like profile for the broad NaID emission and absorption (see the text for additional details).
}\label{f:naid_fitting_explained}
\end{figure*}

Next, we model the various NaID components. As noted previously, we find a surprising similarity between the kinematics of the narrow and broad NaID and the kinematics of the narrow and broad ionized emission lines. If the neutral and the ionized phases are indeed connected in our source, we can use the best-fitting kinematic parameters of the ionized emission lines to constrain the kinematics of the NaID absorption and emission. We visualize this process in figure \ref{f:naid_fitting_explained}, where the left panel shows a best-fitting profile for the narrow and broad ionized emission, and the right panel shows the NaID emission and absorption templates built using the ionized emission kinematics. The full NaID profile is a combination of two such templates, separated by the appropriate wavelength difference of the two NaID doublet lines.

The narrow NaID emission, which is marked with a green line in figure \ref{f:naid_fitting_explained}, is given by:
\begin{equation}\label{eq:2}
	\begin{split}
	& \mathrm{f_{narrow\,NaID}(\lambda) = A_{n,K} e^{-(\lambda - \lambda_{n,K})^2 / 2 \sigma_{n}^2} +    } \\
	& \mathrm{ \,\,\, + 2 A_{n,K} e^{-(\lambda - \lambda_{n,H})^2 / 2 \sigma_{n}^2}},
	\end{split}
\end{equation}
where $\lambda_{n,K}$ and $\lambda_{n,H}$ are the central wavelengths of the narrow NaID $K$ and $H$ components, which are determined from the narrow ionized emission line velocity with respect to systemic. $\mathrm{\sigma_{n}}$ is the velocity dispersion of the narrow emission lines, which is the best-fitting velocity dispersion of the narrow ionized lines\footnote{The assumption that the NaID emission shares similar kinematics with the H$\alpha$ emission suggests that the cool neutral gas and warm ionized gas share the same kinematics. While this is clearly what we observe in this particular system, it is certainly not a general statement, and different gas phases can show very different kinematics (see e.g., \citealt{rupke05c, rupke13, cazzoli16, rupke17}).}. The intensity ratio of the two NaID doublet lines can range between 1 and 2, corresponding to optically-thick and optically-thin gas respectively. Throughout our experiments, when we let the intensity ratio to vary during the fit, we found best-fitting values that are close to 2. We therefore force the intensity ratio of the two lines to be 2.  Thus, our model of the narrow NaID emission, $\mathrm{f_{narrow\,NaID}(\lambda)}$, has only one free parameter, which is the amplitude of the NaID~$\lambda$ 5897\AA\, component, $\mathrm{A_{n,K}}$. We note that the narrow NaID emission shows two well-resolved doublet lines, and thus this component is not significantly degenerate with the broad NaID emission we describe below.

To model the broad NaID emission and absorption, we use the best-fitting broad emission line profile, which includes contributions from both the receding and approaching sides of the outflow, and convert it into a template P-Cygni-like profile for the NaID absorption and emission. The broad NaID absorption and emission templates are constructed by splitting the broad emission line into two parts using some velocity threshold value, marked with red in figure \ref{f:naid_fitting_explained}. The redshifted part of the emission line (with respect to the velocity threshold) serves as the broad redshifted NaID emission template, and both are marked with a blue line in figure \ref{f:naid_fitting_explained}. The blueshifted part of the emission line (with respect to the \emph{same} velocity threshold) serves as the broad blueshifted NaID absorption template, and both are marked with a cyan line in the figure  \ref{f:naid_fitting_explained}. The velocity threshold used for the splitting is, in principle, a free parameter of the model. Throughout our experiments we found that the best-fitting velocity threshold is close to the velocity of the narrow emission. We therefore force the velocity threshold to be equal to the velocity of the narrow emission, which is given by the best-fitting velocity of the narrow H$\alpha$ emission line. 

\begin{figure*}
\includegraphics[width=1\textwidth]{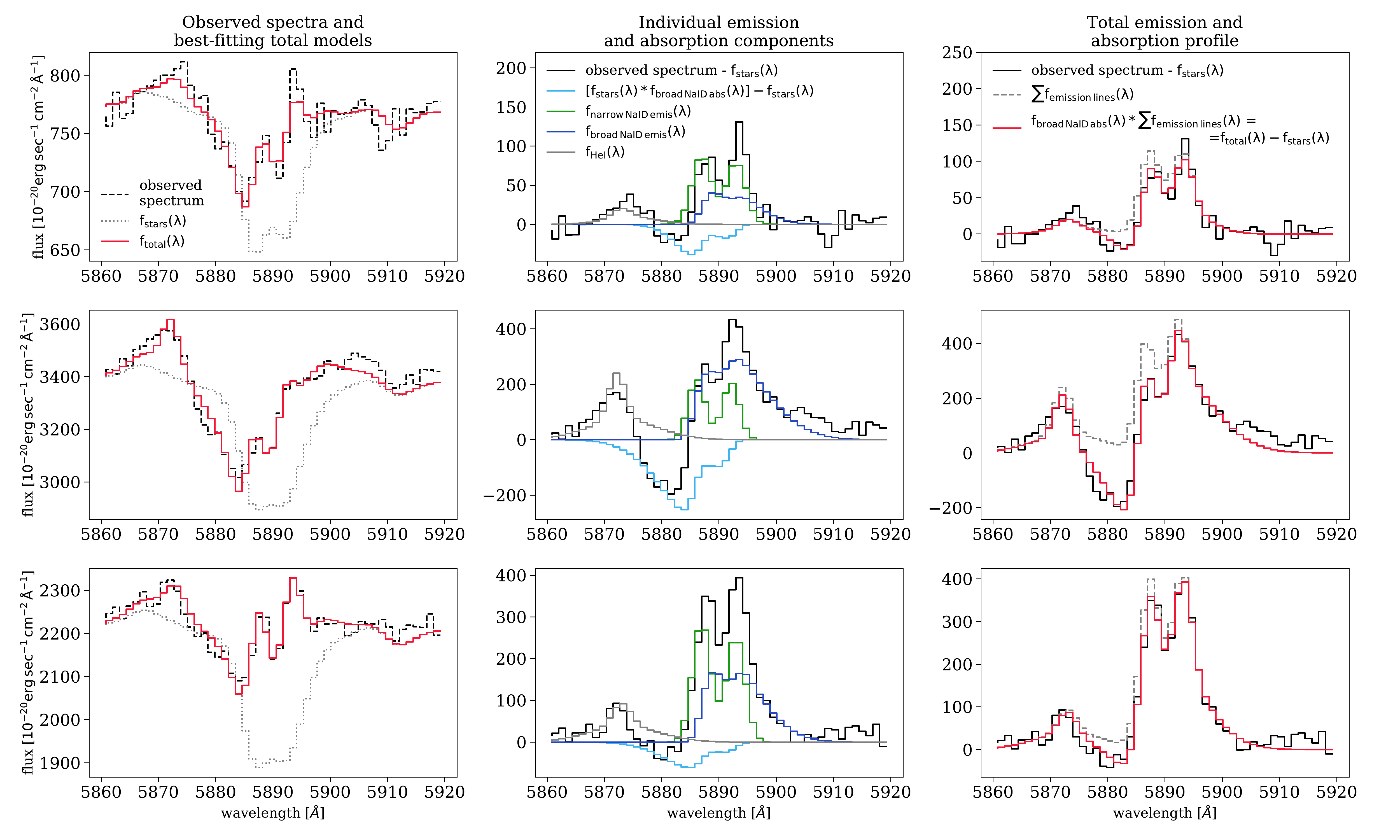}
\caption{\textbf{Three representative spectra around the NaID region and their best-fitting models according to equation \ref{eq:1}.} Each row represents one spaxel in the FOV. The first column shows the observed spectrum (dashed black line), the best-fitting total model $\mathrm{f_{total}(\lambda)}$ (red line), and the stellar model $\mathrm{f_{stars(\lambda)}}$ (grey dotted line) of this spaxel. To illustrate the individual best-fitting emission and absorption components, it is more convenient to subtract the stellar model from the observed spectrum. The second column shows observed spectrum after the stellar model subtraction (black solid line), and presents the individual best-fitting emission and absorption components: narrow NaID emission (green solid line), broad NaID emission (blue solid line), narrow and broad HeI emission (grey solid line), and broad NaID absorption (cyan solid line). The third column represents an integration of the different components shown in the second column, where the observed spectrum after the stellar model subtraction is marked with a black solid line. The contribution of all the emission lines is marked with a grey dotted line, and the contribution of all emission and absorption, which is essentially the best-fitting model $\mathrm{f_{total}(\lambda) - f_{stars(\lambda)}}$, is marked with a red solid line.
}\label{f:emis_and_abs_fitting_example_neutral_only_with_explanations}
\end{figure*}

The broad NaID emission, which is marked with a blue line in figure \ref{f:naid_fitting_explained}, is given by:
\begin{equation}\label{eq:3}
	\begin{split}
	& \mathrm{f_{broad\,NaID\,emis}(\lambda)} =
    	\begin{cases}
   		\mathrm{A_{b,K} e^{-(\lambda - \lambda_{b,K})^2 / 2 \sigma_{b}^2}} & \lambda > \lambda_{n,K},  \\
    	0 & \mathrm{otherwise}.
    	\end{cases}
	+ \\
    	\,\,\,\, & \begin{cases}
   		\mathrm{2 A_{b,K} e^{-(\lambda - \lambda_{b,H})^2 / 2 \sigma_{b}^2}} & \lambda > \lambda_{n, H},  \\
    	0 & \mathrm{otherwise}.
    	\end{cases}
	\end{split},
\end{equation}
where $\lambda_{b,K}$ and $\lambda_{b,H}$ are the central wavelengths of the broad NaID components, which are determined from the broad emission line velocities, and $\lambda_{n,K}$ and $\lambda_{n,H}$ are the central wavelengths of the narrow NaID components, which are used as the velocity threshold. $\mathrm{\sigma_{b}}$ is the best-fitting velocity dispersion of the broad ionized lines, and $\mathrm{A_{b,K}}$ is the amplitude of the $\lambda_{K}=5897$ \AA\, Gaussian, which is a free parameter of the model. 

The broad NaID absorption is modeled as part of the P-Cygni profile we constructed, with the same velocity threshold we used for the broad NaID emission. We assume a Gaussian optical depth, which is a common choice when modeling the NaID absorption (\citealt{rupke05a, rupke15, perna19}). Therefore, the broad NaID absorption, which is marked with a cyan line in figure \ref{f:naid_fitting_explained}, is given by:
\begin{equation}\label{eq:4}
	\mathrm{f_{broad\,NaID\,abs}(\lambda)} = \mathrm{e^{-\tau_{b, K}(\lambda) - \tau_{b, H}(\lambda)}}, 
\end{equation}
and the optical depths are given by:
\begin{equation}\label{eq:5}
	\mathrm{\tau_{b, K}(\lambda)} = 
	\begin{cases}
	\mathrm{\tau_{0,K} e^{-(\lambda - \lambda_{b,K})^2 / 2 \sigma_{b}^2}} & \lambda < \lambda_{n,K},  \\	
	0 & \mathrm{otherwise}	
	\end{cases}
\end{equation} and: 
\begin{equation}\label{eq:6}
	\mathrm{\tau_{b, H}(\lambda)} = 
	\begin{cases}
	\mathrm{2 \tau_{0,K} e^{-(\lambda - \lambda_{b,H})^2 / 2 \sigma_{b}^2}} & \lambda < \lambda_{n,H},  \\	
	0 & \mathrm{otherwise}	
	\end{cases},
\end{equation}
where $\mathrm{\tau_{0,K}}$ is the absorption optical depth of the $\lambda_{K}=5897$ \AA\, component, which is a free parameter of the model. Here too, we forced the ratio of optical depths at line center of the two NaID components to be 2. By doing so, we assume that the absorbing gas is optically-thin. This is justified by: (1) when we allow the ratio to vary, the best-fitting ratio is close to 2, and (2) the best-fitting absorption optical depths, $\mathrm{\tau_{0,K}}$ and $\mathrm{\tau_{0,H}}$, are always smaller than 1.

Several studies modeled the NaID absorption as a combination of clouds with different optical depths and different covering factors (see e.g., \citealt{rupke05a, rupke15, perna19}). We examined a more general model for the blueshifted NaID absorption, where we allowed the covering factor to vary, and found that the best-fitting covering factor is always close to 1. In addition, in section \ref{s:derived_gas_props} below we examine the physical properties of the neutral gas, and show that the combination of NaID emission and absorption suggests a covering factor that is close to 1. Therefore, our final adopted model has only three free parameters: $\mathrm{A_{n,K}}$, $\mathrm{A_{b,K}}$, and $\mathrm{\tau_{0,K}}$. In figure \ref{f:emis_and_abs_fitting_example_neutral_only_with_explanations} we show the best-fitting models for three representative spectra, where we break the model into all the individual components for clarity. Despite the few free parameters, the model tracks the observed spectra well. 

Our modeling of the NaID complex is not fully consistent. First, in a classical P-Cygni profile, the transition between the redshifted emission and the blueshifted absorption is expected to be gradual, while we have modeled it using step functions. This results in zero absorption (emission) above (below) a certain velocity threshold. This is a direct result of our model of the broad ionized emission lines, where we use a single broad Gaussian component to describe both the receding and the approaching sides of the outflow (and not two separate kinematic components). This forces us to construct a P-Cygni-like profile for the NaID lines by splitting the broad ionized emission with step functions. A more complete approach would be to model the approaching and receding parts of the ionized outflow by two Gaussians, and use each of the components to model the broad NaID absorption and emission separately. Such a decomposition is not possible for our source. Second, we suggest (see section \ref{s:origin_of_narrow_lines}) that the narrow redshifted NaID emission originates in the receding part of the outflow as well. A symmetric outflow would require an additional narrow blueshifted NaID absorption. We found no evidence for such a component, and thus did not include it in our model. Such a component may exist, but its contribution is negligible compared with the broad blueshifted NaID absorption.

\subsubsection{Derived ionized and neutral gas properties}\label{s:derived_gas_props}

\begin{figure*}
\includegraphics[width=0.99\textwidth]{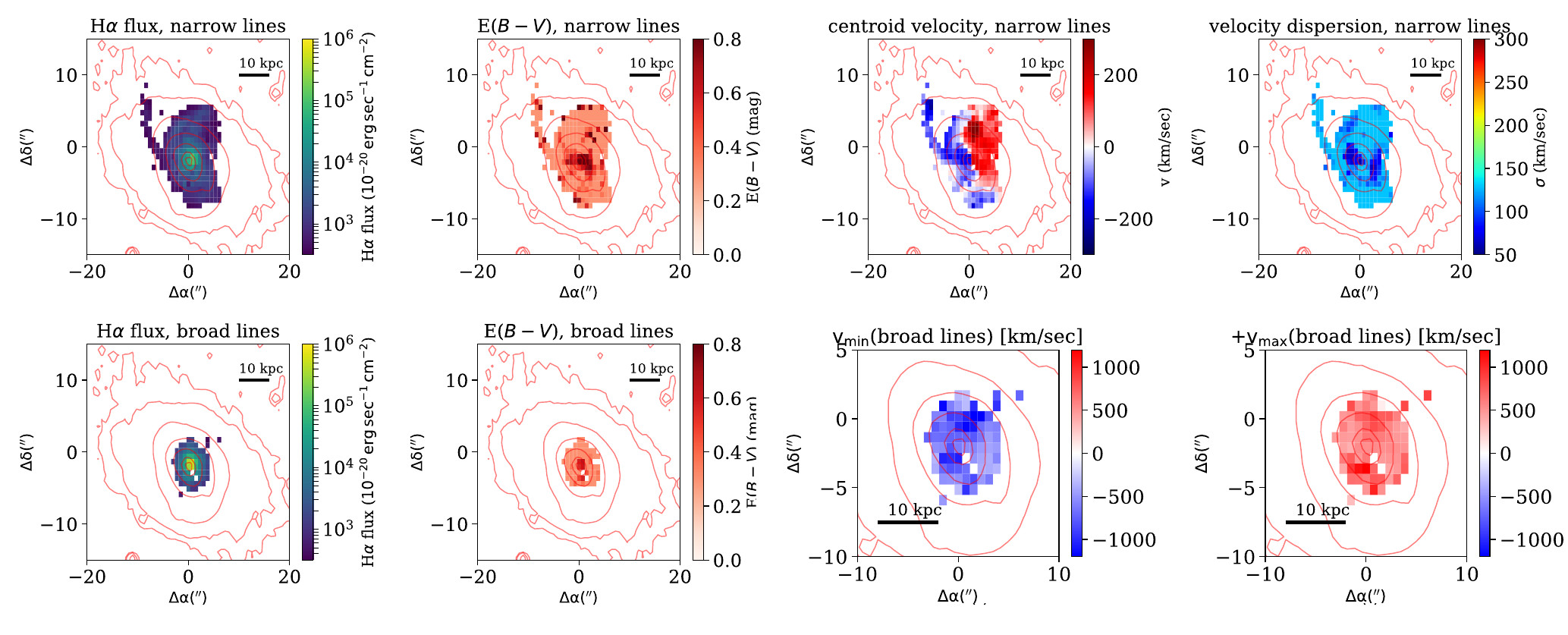}
\caption{\textbf{Derived ionized gas properties in SDSS J124754.95-033738.6.} The top row represents the narrow emission line properties: H$\alpha$ flux, dust reddening towards the emission lines, centroid velocity with respect to systemic velocity, and velocity dispersion. The bottom row represents the broad emission line properties: H$\alpha$ flux, dust reddening towards the emission lines, and the minimal and maximal velocity in each spaxel, which are defined as $\Delta v - 2\sigma$ and $\Delta v + 2\sigma$ respectively. The red contours represent steps of 0.5 dex in stellar continuum emission (see figure \ref{f:2d_stellar_properties}). The black bar represents a distance of 10 kpc at the redshift of the system. While the narrow lines are distributed over tens of kpc, and their centroid follows roughly the stellar rotation pattern, the broad lines are detected up to 5 kpc and show very large positive and negative velocities, suggestive of a fast outflow.
}\label{f:ionized_line_properties_extended}
\end{figure*}

Having decomposed the ionized and neutral emission and absorption, we can now examine the physical properties of the neutral and ionized gas. We start by presenting the properties of the ionized gas. In particular, we examine its spatial distribution and kinematics, its main source of photoionization, the dust reddening, and the electron density in the outflow. We then present the derived properties of the neutral gas, such as emission line luminosity, absorption EW, optical depth and covering factor. Finally, we discuss the observed connection between the NaID absorption and emission in this system.

In figure \ref{f:ionized_line_properties_extended} we present several properties of the narrow and broad ionized emission lines. For the narrow lines, we show the integrated H$\alpha$ flux, the dust reddening towards the emission lines (see equation 1 in \citealt{baron19b}), the centroid velocity with respect to systemic velocity, and the velocity dispersion. For the broad lines, we show the H$\alpha$ flux, the dust reddening towards the emission lines, and minimal and maximal outflow velocities, which are defined as $\Delta v - 2\sigma$ and $\Delta v + 2\sigma$ respectively. Similar to figure \ref{f:ionized_gas_emission}, the narrow H$\alpha$ flux appears to be disturbed and asymmetric, and it extends to distances larger than 30 kpc. The broad H$\alpha$ flux appears to be more symmetric, and it extends to distances of 5 kpc from the center of the galaxy. Similarly to other post starburst E+A galaxies with ionized outflows (\citealt{baron17b, baron18}), the dust reddening towards the narrow and broad lines is high, $\mathrm{E}_{B-V} \sim 0.6$ mag, which is higher than the dust reddening typically observed in type II AGN that host ionized outflows \citep{baron19b}.

The centroid velocity of the narrow lines is consistent with rotation, and appears to be roughly similar to the rotation pattern of the stars (see figure \ref{f:2d_stellar_properties}). We find approximately the same rotational axis for the narrow emission lines and the stars, but we find some inconsistency in the velocity field at a distance of 6--10 kpc in the north-south (up-down) direction. This might suggest that some of the narrow emission lines are not emitted by stationary NLR-like gas, but rather by the outflowing gas, which we further discuss in section \ref{s:origin_of_narrow_lines}. The velocity dispersion of the narrow lines is rather low and uniform throughout the FOV, except for two regions with a somewhat lower velocity dispersion, both of which coincide with an upturn of a spiral arm. 

Since we fit a single broad Gaussian to describe both the approaching and the receding sides of the outflow, the centroid velocity and velocity dispersion of the broad lines are difficult to interpret. Instead, it is useful to examine the minimum and maximum velocity in each spaxel. The maximal positive and negative velocities are quite similar in all spaxels: $\sim \pm$1\,000 km/sec. The symmetry between the positive and negative maximal velocities suggests that if this is indeed a double-cone, it is in a nearly face-on configuration. An alternative scenario is a full expanding shell ,where the opening angle is close to $90^{\circ}$. However, in such a scenario, one would expect to find lower velocities in spaxels at the edges of the outflow due to projection effects. We find no evidence for this. In section \ref{s:outflow_geometry} we discuss the geometry of the outflow in detail, and summarize the evidence for a face-on double-cone outflow.

\begin{figure*}
\includegraphics[width=1\textwidth]{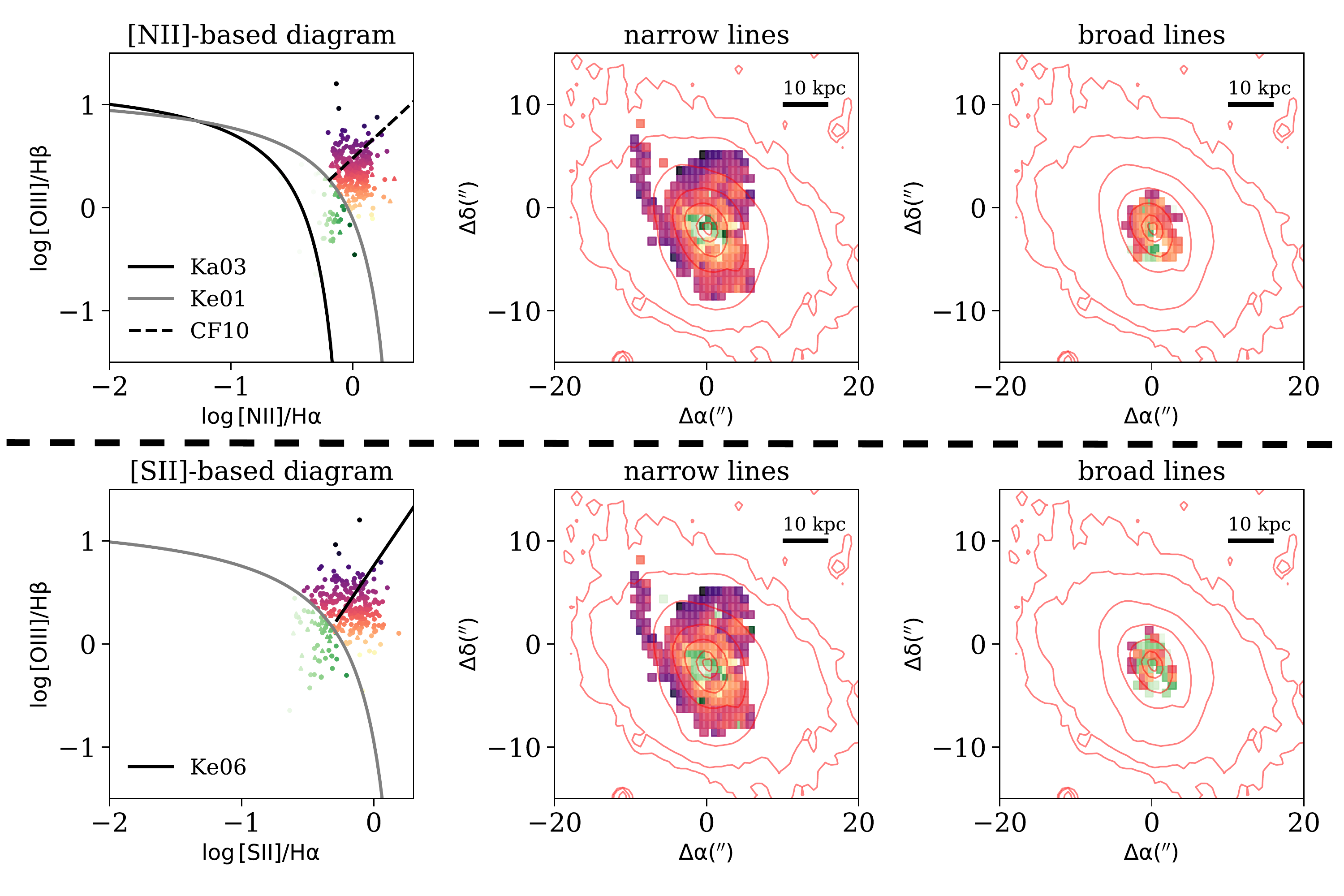}
\caption{\textbf{Line diagnostic diagrams showing the main source of photoionization in the warm ionized gas, and its degree of ionization.} The left column shows the emission line ratios $\mathrm{\log{} [OIII]/H\beta}$ versus $\mathrm{\log{} [NII]/H\alpha}$ (top) and $\mathrm{\log{} [OIII]/H\beta}$ versus $\mathrm{\log{} [SII]/H\alpha}$ (bottom). We mark the two separating criteria that are used to separate star forming from AGN-dominated galaxies (\citealt{kewley01, kauff03a}; Ke01 and Ka03 respectively), and the two LINER-Seyfert separating lines from \citet[CF10]{cidfernandes10} and \citet[Ke06]{kewley06}. We mark the narrow lines with circles and the broad lines with triangles. The middle and right columns show the spaxels in which we detected narrow (middle column) and broad (right column) kinematic components, color-coded according to their location in the line-diagnostic diagrams. The narrow and broad kinematic components are consistent with AGN photoionization in the large majority of spaxels, with some contribution from HII regions in the central regions of the galaxy. The narrow $\mathrm{[OIII]/H\beta}$ line ratio suggests that the ionization of the gas increases with increasing distance. The more-or-less constant broad $\mathrm{[OIII]/H\beta}$ line ratio suggests a constant ionization parameter and hence gas which is at the same distance from the central source. }\label{f:emission_line_ratios_new}
\end{figure*}

Next, we examine the main source of ionizing radiation in the regions emitting the narrow and broad emission lines, as well as their degree of ionization. In figure \ref{f:emission_line_ratios_new} we show the narrow and the broad emission components on several standard line diagnostic diagrams. The narrow and broad kinematic components are consistent with AGN photoionization in the large majority of spaxels, with some contribution from HII regions in the central regions of the galaxy. We estimate the relative contribution of SF to the ionization of the narrow and broad emission lines, using the prescription by \citet{wild10}. Outside the few central spaxels, we find a negligible contribution of SF to the ionization of the narrow lines, with a relative contribution of 0--0.2. For the broad lines, we find a contribution of roughly 0.3. We therefore conclude that the AGN is the main source of ionizing radiation of both narrow and broad kinematic components in the majority of the spaxels, with a significant contribution from SF only in the central region.

The degree of ionization of the gas can be roughly estimated using the observed $\mathrm{[OIII]/H\beta}$ line ratio, since this ratio strongly depends on the ionization parameter of the ionized gas (see \citealt{baron19b} and references therein). In particular, for the ionization parameters observed in this system, we expect the $\mathrm{[OIII]/H\beta}$ line ratio to be larger in more ionized regions, and to be smaller in less ionized regions. Figure \ref{f:emission_line_ratios_new} shows a continuous change in the \emph{narrow} $\mathrm{[OIII]/H\beta}$ line ratio as a function of distance from the center of the galaxy, with smaller $\mathrm{[OIII]/H\beta}$ ratios close to the center and larger ratios farther out. This suggests that there is a continuous change in the degree of ionization of the line emitting gas, with lower ionization closer to the center of the galaxy and higher ionization farther out. This suggests that the density falls faster than $\mathrm{n_{H}(r) \propto r^{-2}}$. In contrast to the narrow lines, the \emph{broad} $\mathrm{[OIII]/H\beta}$ line ratio remains nearly constant and consistent with LINER-like emission throughout the FOV, suggesting that the degree of ionization of the gas that emits the broad lines is roughly the same in the different spaxels. Since the degree of ionization depends on the gas density and on its distance from the central ionizing source, a constant gas density would imply that the gas that emits the broad lines is located at the same distance from the central source. Finally, we estimated the ionization parameter in the narrow and broad kinematic components using the expression from \citet{baron19b}. The ionization parameter of the narrow line-emitting gas ranges from $\log{U} = -3.7$ in the center to $\log{U} = -3.2$ at 15 kpc. The ionization parameter of the broad line-emitting gas is roughly constant throughout the FOV at $\log{U} = -3.7$. These estimates are used in the photoionization models in section \ref{s:photoionization_models}.

An alternative scenario to account for the radial variation in the narrow [OIII]/H$\beta$ is mixing of SF and AGN ionization throughout the FOV, where regions that are dominated by SF show lower [OIII]/H$\beta$ ratios, and regions dominated by the AGN showing larger [OIII]/H$\beta$ ratios. We find this scenario to be unlikely since: (1) the relative contribution of SF to the ionization of the narrow lines is negligible outside the central region, and (2) a varying contribution of SF and AGN to the gas ionization results in diagonal "mixing tracks" on the BPT diagram, where both [OIII]/H$\beta$ and [NII]/H$\alpha$ line ratios increase with increasing AGN contribution (see e.g., \citealt{wild10, davies14}). In our source we find a vertical trend, where the [OIII]/H$\beta$ varies dramatically while the [NII]/H$\alpha$ remains roughly constant. This is in line with the photoionization models of \citet{baron19b}, that show that for gas that is photoionized by an AGN, a variation of the ionization parameter results in a vertical trend of increasing [OIII]/H$\beta$ ratio and quite constant [NII]/H$\alpha$ ratio. We therefore conclude that the radial variation in the narrow [OIII]/H$\beta$ ratio is most likely due to a variation of the ionization parameter. 

\begin{figure*}
\includegraphics[width=0.9\textwidth]{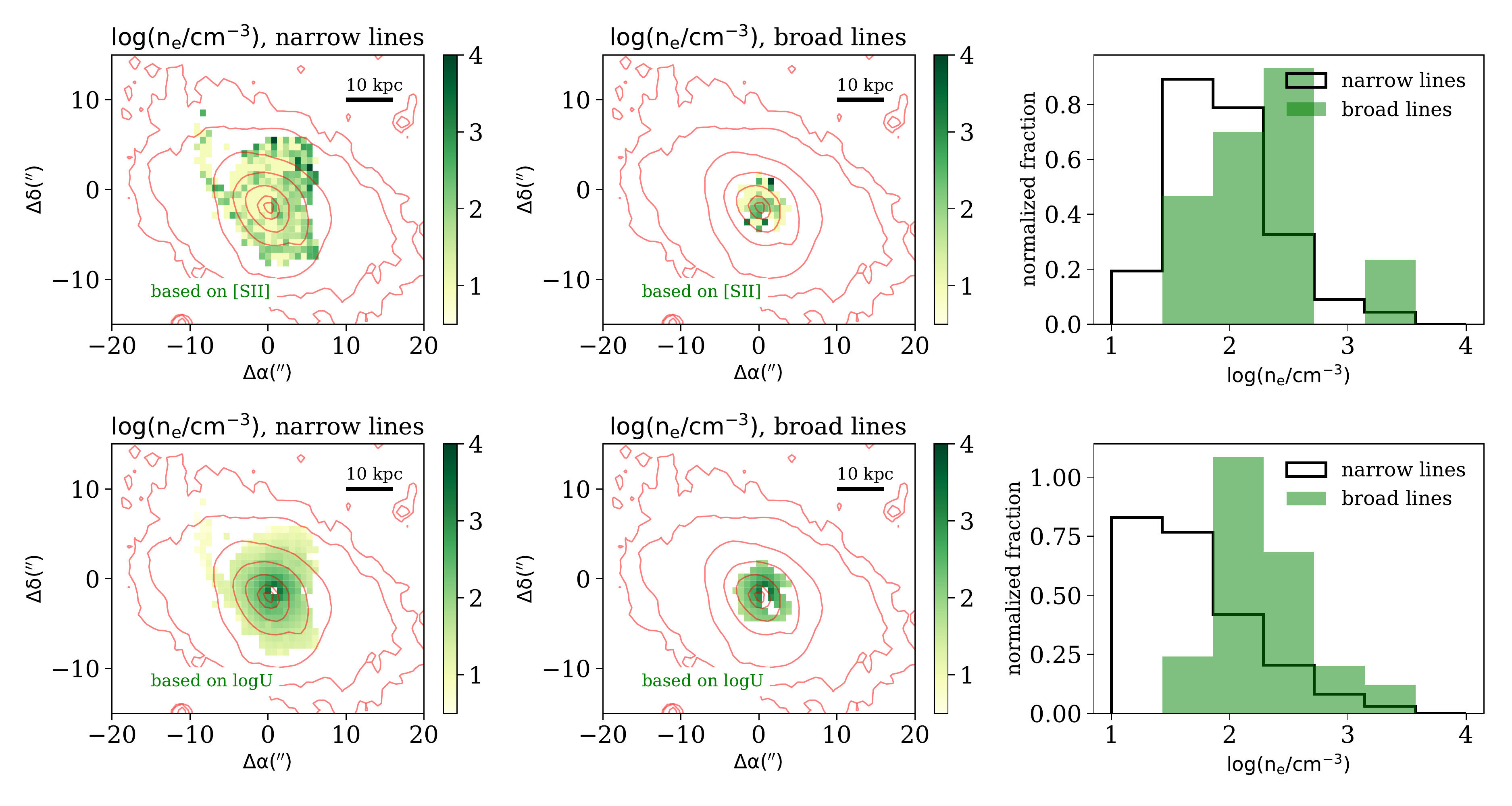}
\caption{\textbf{Two methods to measure the electron density in the warm ionized gas.} The first is based on the commonly-used [SII] line ratio (top panels), and the second on the $\mathrm{\log{} [OIII]/H\beta}$ and $\mathrm{\log{} [NII]/H\alpha}$ line ratios, and on the known distance of gas from the central source (bottom panels). The left panels show the spatial distribution of the electron densities measured for the narrow lines, the middle panels show the spatial distribution for the broad lines, and the right panels compare the histograms of electron densities of both of the kinematic components. The two methods give consistent electron densities throughout the FOV. This suggests that the projected distance is similar to the actual distance of the gas, which indicates that the opening angle of the outflowing cone is not small ($\gtrsim 45^{\circ}$; see text for additional details). }\label{f:electron_densities_all}
\end{figure*}

Next, we estimate the electron densities in the warm ionized gas using two methods. The first method is based on the commonly-used [SII] line ratio and is sensitive to electron densities in the range $10^{2}\, \mathrm{cm^{-3}} < n_{e} < 10^{4} \, \mathrm{cm^{-3}}$ (e.g., \citealt{osterbrock06}). The top panels in figure \ref{f:electron_densities_all} show the results for this method, where we show the spatial distribution of electron densities in the narrow line-emitting gas in the left panel, the distribution in the broad line-emitting gas in the middle panel, and compare the electron density histograms of the two components in the right panel. In \citet{baron19b} we discussed the uncertainties that are associated with electron densities based on the [SII] method. In particular, we showed that the ratio is poorly constrained for the majority of the objects in our type II sample, and the resulting electron densities were consistent with the entire range of possible electron densities. For SDSS J124754.95-033738.6, the exceptional quality of the MUSE observations allows us to place tight constraints on the [SII] line ratios in most of the spaxels, resulting in robust electron density estimates. 

\citet{baron19b} proposed a novel method to estimate the electron density in ionized gas, which is based on the bolometric luminosity of the AGN, the optical $\mathrm{[OIII]/H\beta}$ and $\mathrm{[NII]/H\alpha}$ line ratios, and the distance of the gas from the central source (see sections 4.2 and 5.2.2 in \citealt{baron19b}). The electron density is given by:
\begin{equation}\label{eq:7}
	{n_{\mathrm{e}} \approx 3.2 \Big(\frac{L_{\mathrm{bol}}}{10^{45}\, \mathrm{erg/sec}}\Big) \Big( \frac{r}{1\,\mathrm{kpc}} \Big)^{-2} \Big(\frac{1}{U}\Big) \, \mathrm{cm^{-3}}       },
\end{equation}
where $L_{\mathrm{bol}}$ is the AGN bolometric luminosity and $r$ is the distance from the central source. The ionization parameter, $U$, is given by (\citealt{baron19b}):
\begin{equation}\label{eq:8}
	\begin{split}
& \log U = a_{1} + a_{2}\Big[\log\mathrm{\Big(\frac{[OIII]}{H\beta}\Big)}\Big] + a_{3}\big[\log\mathrm{\Big(\frac{[OIII]}{H\beta}\Big)}\Big]^{2} + \\
& + a_{4}\Big[\log\mathrm{\Big(\frac{[NII]}{H\alpha}\Big)}\Big] + a_{5}\Big[\log\mathrm{\Big(\frac{[NII]}{H\alpha}\Big)}\Big]^{2},
	\end{split}
\end{equation}
where the constants ($a_{1}$, $a_{2}$, $a_{3}$, $a_{4}$, $a_{5}$) are given by (-3.766, 0.191, 0.778, -0.251, 0.342). In our case, we only know the projected distance, which can be very different from the actual distance. In particular, if the broad emission lines originate from a double cone outflow that is viewed face on, the difference between the projected distance and the actual distance is related to the opening angle of the cones, such that for small opening angles, which implies elongated cones, the projected distance can be much smaller than the actual distance. For a large opening angle, the projected distance is roughly similar to the actual distance. 

Since the ionization parameter-based method depends on the location of the outflow while the [SII]-based method does not, we can estimate the de-projected distance of the gas by comparing the densities derived with the two methods. We estimate the electron densities using equations \ref{eq:7} and \ref{eq:8}, taking the distance to be the measured projected distance. If the opening angle of the cone is large, and the projected distance is roughly similar to the actual distance, we expect to find similar electron densities using the two methods: $\mathrm{n_{e}([SII]) = n_{e}(U;r=r_{projected})}$. On the other hand, if the opening angle is small, and the projected distance is much smaller than the actual distance, we expect the electron densities estimated using our method to be larger than those estimated using the [SII]-method. For example, for an opening angle of 6$^{\circ}$, the projected distance is ten times smaller than the actual distance, and we would expect a factor of 100 difference between the two electron density estimates. For example, for $\mathrm{n_{e}([SII]) = 10^{2}\, cm^{-3}}$, the estimated electron density using the ionization parameter method will be $\mathrm{n_{e}(U; r=r_{projected}) = 10^{4}\, cm^{-3}}$. 

\citet{baron19b} noted that the [SII]-based electron densities can be lower than the [OIII] and H$\alpha$-based densities because the [OIII] and H$\alpha$ lines are emitted throughout most of the ionized cloud, while the [SII] lines are emitted close to the ionization front, where the electron density can be significantly smaller. This effect can account for a difference in densities of up to a factor of 10, but not larger. Due to the degeneracy between the two scenarios (small opening angle versus the photoionization considerations mentioned above), this comparison cannot be used to accurately determine the de-projected distance of the gas. However, it can be used to check whether the de-projected distance is close to the observed projected one, as we do here.

In the bottom panels of figure \ref{f:electron_densities_all} we show the electron densities estimated using our method. One can see that the electron densities inferred for the narrow lines are roughly consistent throughout the FOV. Nevertheless, we do find a significant difference between the electron density estimates in the east (right) side of the narrow line-emitting region. Inspection of the best-fitting spectra reveals that the SNR of the [SII] lines is small in these regions, which might explain the discrepancy. We find that the electron densities inferred for the broad lines are of the same order of magnitude for the two different methods. This suggests that the projected distance is not very different from the actual distance, and that the opening angle of the cone is not small ($\gtrsim 45^{\circ}$, where $90^{\circ}$ represents a full shell). We therefore adopt $r=5\,\mathrm{kpc}$, which is the projected distance of the spaxels located on the edge of the broad line-emitting region, as the representative distance of the outflowing gas from the central source.

The electron densities we infer for the ionized outflow in this object are about $n_{e} \sim 10^{2}\,\mathrm{cm^{-3}}$, which is two orders of magnitude lower than our estimated electron densities in ionized outflows in type II AGN \citep{baron19b}. This difference can be attributed to: (1) SDSS J124754.95-033738.6 is a post starburst E+A galaxy, while the type II AGN sample studied in \citet{baron19b} consists of typical star forming galaxies, and (2) we estimate the location of the ionized outflow in SDSS J124754.95-033738.6 to be $r \sim$5 kpc, while the typical location of the outflows in the type II AGN sample is much smaller, $r \sim$ 200 pc (\citealt{baron19a}). 

\begin{figure*}
	\centering
\includegraphics[width=0.7\textwidth]{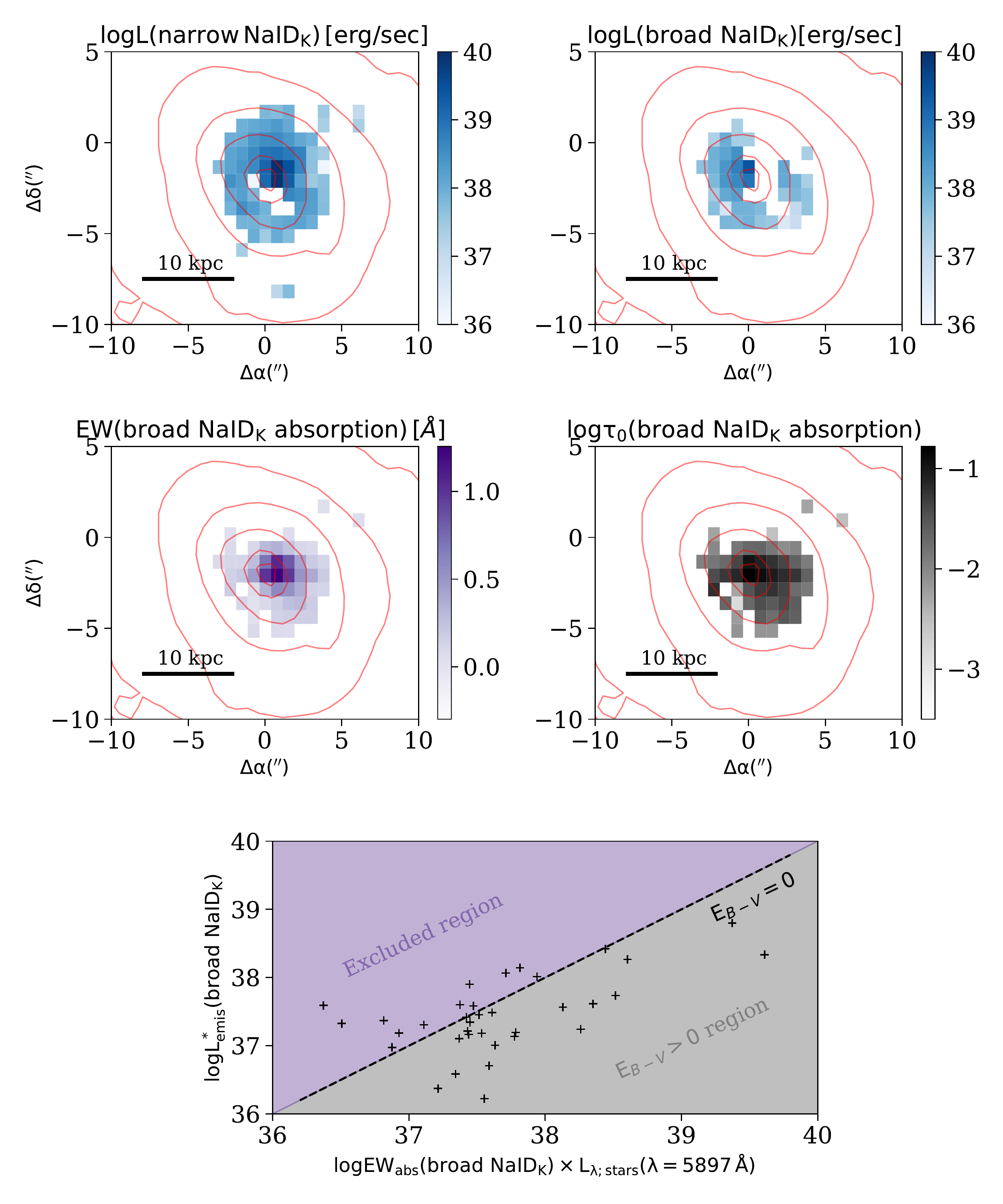}
\caption{\textbf{Properties of the NaID emission and absorption lines.} \textbf{The first row} represents the emission properties, where we show the dust-corrected NaID$_{K}$ luminosities for the narrow and broad kinematic components (left and right panels respectively). \textbf{The second row} represents the absorption properties, where we show the EW and optical depth at line center of the broad NaID$_{K}$ component (left and right panels respectively). The insets of the first two rows summarize the fitted kinematic components and the chosen constrains during the fit. \textbf{The third row} presents the observed connection between the broad NaID emission and absorption. The panel shows L$^{*}_{\mathrm{emis}}$(broad NaID$_{K}$), which is the NaID$_{K}$ luminosity prior to dust correction, versus $\mathrm{ EW_{abs}(broad\,\, NaID_{K}) \times L_{\lambda; \, stars}(\lambda=5897\,\AA) }$, where $\mathrm{EW_{abs}(broad\,\, NaID_{K})}$ is the absorption EW, and $\mathrm{L_{\lambda; \, stars}(\lambda=5897\,\AA)}$ is the stellar continuum at the NaID$_{K}$ wavelength. The dashed grey line represents the expected relation between the absorption and emission for a symmetric neutral wind without dust. Dusty outflow are expected to occupy the region beneath the dashed line (grey background color). The area above the line is excluded for symmetric outflows (purple background color). The distribution in the different spaxels suggests that the wind is dusty and its approaching and receding sides are roughly symmetrical. }\label{f:naid_properties_all_with_explanations}
\end{figure*}

Finally, we use the best-fitting models of the NaID complex to study the neutral gas properties. In figure \ref{f:naid_properties_all_with_explanations} we present a summary of the NaID emission and absorption properties. The first row shows the dust-corrected narrow and broad NaID emission line luminosities, where we used the reddening derived from the narrow and broad ionized emission lines respectively. This correction is justified since according to our proposed model, which we further discuss in section \ref{s:photoionization_models}, the NaID and the ionized emission lines are emitted by the same clouds, and therefore we expect them to be obscured by roughly the same dust column densities. The second row represents the broad NaID absorption properties, where we show the EW and optical depth at line center for the NaID$_{K}$ component. The optical depth is low, $\tau_{0} < 0.1$, suggesting that the absorbing gas is optically-thin. The NaID emission and absorption coincide in most of the spaxels throughout the FOV, with stronger emission and absorption closer to the center of the galaxy, and weaker emission and absorption farther out. This reinforces our suggestion for a face-on outflow that produces a P-Cygni-like profile.

According to our model, the broad NaID emission and absorption originate from a double-cone neutral outflow, where the redshifted emission is due to the receding part of the outflow, while the blueshifted absorption is due to the approaching side. We can therefore compare the emission and absorption properties and examine how symmetric the outflow is. If the approaching and receding sides of the outflow are completely symmetric (same density, spatial extent, covering factor, etc), and assuming no dust, we expect the blueshifted absorption and redshifted emission to have the same EW (NaID is a resonant transition, and every absorption results in reemission). Therefore, the NaID emission line luminosity can be expressed as: $\mathrm{L_{emis}(NaID) = EW_{abs}(NaID) \times L_{\lambda; stars}(NaID)}$, where $\mathrm{EW_{abs}(NaID)}$ is the absorption EW, and $\mathrm{L_{\lambda; stars}(NaID)}$ is the stellar continuum at the NaID wavelength. For dusty gas, the EW of the redshifted emission will be smaller than the blueshifted absorption, since photons that are emitted by the receding side of the outflow go through a larger column density of dust. As a result, $\mathrm{L_{emis}(NaID) < EW_{abs}(NaID) \times L_{\lambda; stars}(NaID)}$. We show these two scenarios in the bottom row of figure \ref{f:naid_properties_all_with_explanations}, where a symmetric outflow with no dust is marked with a dashed line. Dusty outflows will occupy the region beneath the dashed line (grey color). In case of a symmetric outflow, the area above the line (purple color) is excluded. 

We compare the broad NaID emission to the broad NaID absorption in the bottom rows of figure \ref{f:naid_properties_all_with_explanations}. Since we want to examine the effect of dust reddening, the y-axis shows the broad NaID$_{K}$ luminosity without correcting for dust reddening (hereafter L$^{*}$(NaID$_{K}$)). The x-axis shows $\mathrm{ EW_{abs}(broad\,\, NaID_{K}) \times L_{\lambda; stars}(\lambda=5897\,\AA) }$, where $\mathrm{EW_{abs}(broad\,\, NaID_{K})}$ is the absorption EW, and $\mathrm{L_{\lambda; stars}(\lambda=5897\,\AA)}$ is the stellar continuum at the NaID$_{K}$ wavelength, which we take from the best-fitting stellar model. The product of the two represents the expected NaID$_{K}$ luminosity produced by the approaching side of the outflow (via scattering of the absorption photons), as would have been observed by an observer from the opposite direction. By comparing the luminosities of the receding (x-axis) and approaching (y-axis) sides of the outflow, we can examine how symmetric the outflow is. One can see that most of the spaxels occupy the $\mathrm{E}_{B-V} > 0$ region in the diagram, and only a small minority occupy the "excluded" region in the diagram. Furthermore, if instead we show the \emph{dust-corrected} broad NaID$_{K}$ luminosity in the y-axis, most of the measurements cluster around the $\mathrm{E}_{B-V} = 0$ line. While these arguments are somewhat simplistic and the measurements are subjected to significant uncertainties, they are consistent with the suggestion that the neutral outflow gas is dusty, and that the approaching and receding parts of the outflow are roughly symmetric.

\subsubsection{Connection between the neutral and ionized gas phases}\label{s:connection_between_gas_phases}

\begin{figure*}
\includegraphics[width=0.8\textwidth]{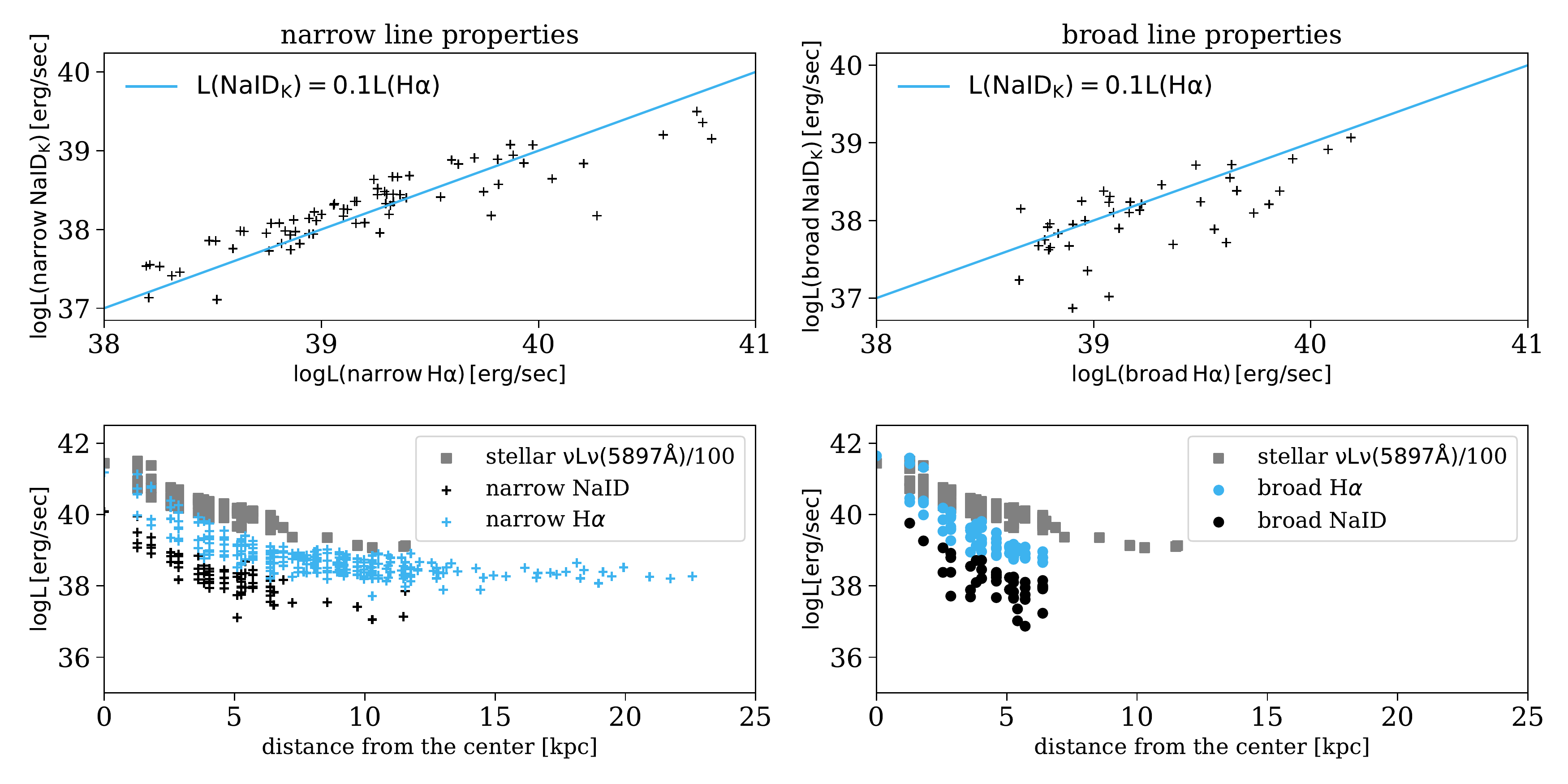}
\caption{\textbf{Connection between the neutral and ionized outflow phases.} The top row compares between the dust-corrected NaID$_{K}$ luminosity and the H$\alpha$ luminosity, where we show the narrow and broad kinematic components in the left and right panels respectively. The solid blue lines represent the case where L(NaID$_{K}$)=0.1L(H$\alpha$). The bottom row compares between the spatial extents of the NaID and H$\alpha$ emission, where we show the narrow and broad kinematic components in the left and right panels respectively. We mark the rescaled stellar continuum luminosity, $\mathrm{\nu L_{\nu}(5897 \AA)/100}$ with grey rectangles, the NaID emission with blue crosses, and the H$\alpha$ emission with black crosses. The narrow and broad kinematic components show a constant and similar L(NaID)/L(H$\alpha$) ratio throughout the FOV. In addition, the NaID and H$\alpha$-emitting gas show similar spatial extents. These observations suggest that the neutral and ionized gas phases originate from the same outflowing clouds. }\label{f:connection_between_the_neutral_and_ionized_phases}
\end{figure*}

In section \ref{s:model_inedpendent_props} we already noted the similarity between the kinematics of the NaID emission and absorption and the kinematics of the ionized emission lines. We used this similarity to model the NaID emission and absorption complex, where we set the kinematic parameters of the different NaID components to be the same as the best-fitting kinematic parameters of the ionized emission lines. The success of these fits in all the spaxels (see e.g., figure \ref{f:emis_and_abs_fitting_example_neutral_only_with_explanations}) reinforces the suggestion of a kinematic connection between the two phases.

In figure \ref{f:connection_between_the_neutral_and_ionized_phases} we further compare the neutral and ionized line luminosities and spatial extents. In the top row we show the dust-corrected NaID and H$\alpha$ luminosities for the narrow and broad kinematic components. The two luminosities show a strong correlation in both cases, with a constant luminosity ratio of L(NaID$_{K}$)/L(H$\alpha$)$\sim 0.1$. This correlation remains as significant when we compare the NaID$_{K}$ and H$\alpha$ luminosities prior to the reddening correction, suggesting that the correlation is not driven by the dust distribution. In the bottom row we examine how these luminosities change as a function of projected distance from the center. The panels show the reddening-corrected narrow and broad NaID$_{K}$ and H$\alpha$ luminosities, and the reddening-corrected stellar continuum at $\lambda = 5897$\AA\, as a function of projected distance from the center of the galaxy. The neutral and ionized emission lines show similar spatial extents in the broad kinematic component throughout the FOV, and similar extents and behavior in the narrow kinematic component up to a distance of 5 kpc. These similarities suggest a common origin of the neutral and ionized emission lines. We use these in our proposed model in section \ref{s:models} below.

\begin{figure*}
\includegraphics[width=1\textwidth]{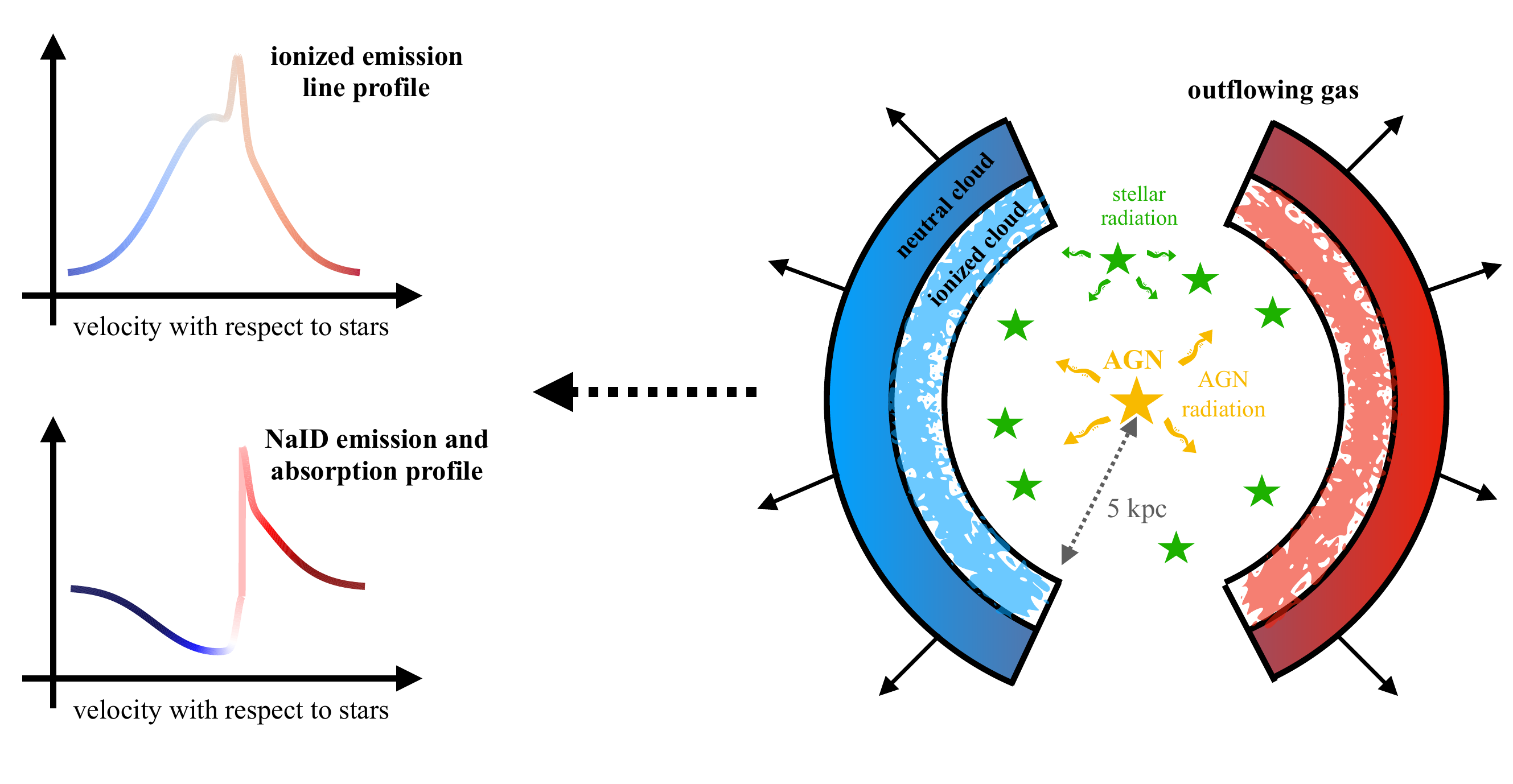}
\caption{\textbf{The proposed outflow geometry.} The ionized emission and neutral emission and absorption originate from the same outflowing clouds, which are embedded in a face-on double-cone outflow with a moderate opening angle ($\sim 45^{\circ}$), at a distance of $\sim$5 kpc from the center. The central AGN ionized the inner parts of the outflowing clouds, creating an ionization structure that produces the observed ionized lines. The observed ionized emission line profile shows a combination of redshifted and blueshifted emission, where the redshifted emission is due to the receding part of the outflow, while the blueshifted is due to the approaching side. The hydrogen column density in these clouds is large enough to make the clouds radiation bounded. As a result, the outer parts of the clouds are neutral, where NaI atoms can absorb and reemit stellar photons. This results in a P-Cygni-like profile for the NaID, where the redshifted NaID emission originates in the receding side of the outflow, and the blueshifted absorption originates in the approaching side. This model naturally explains the observed similarities between the neutral and ionized gas phases. }\label{f:doodle_interpretation}
\end{figure*}

\section{Modeling the outflow}\label{s:models}

In this section we construct a model for the outflow in the system. We propose that the ionized emission, and neutral emission and absorption, originate from the \emph{same} clouds, which are embedded in a double-cone outflow. We present our proposed model in section \ref{s:outflow_geometry}, where we describe all the observational evidence that support the suggested geometry. Then, in section \ref{s:photoionization_models} we construct photoionization models and show that our suggested model is consistent with the various observational constraints. We then use the observed NaID$_{K}$/H$\alpha$ ratio to constrain the Sodium neutral fraction, the size, and the neutral-to-ionized gas mass ratios in the outflowing clouds. In section \ref{s:mass_and_energy} we use the ionized and neutral gas properties to estimate the mass outflow rate and energetics in the ionized and neutral phases of the observed outflow and in section \ref{s:origin_of_narrow_lines} we present evidence that much of the narrow emission line luminosity also originates in the outflow, rather than from a stationary NLR. 

\subsection{Outflow geometry}\label{s:outflow_geometry}

In figure \ref{f:doodle_interpretation} we present an illustration of the suggested outflow geometry. In this model, the system hosts a face-on double-cone outflow, with a moderate opening angle, at a distance of $\sim$5 kpc from the center. The central AGN ionizes the inner parts of the outflowing clouds, creating an ionization structure that produces the observed ionized lines. The emission line profiles are a combination of redshifted and blueshifted emission, where the redshifted emission is due to the receding part of the outflow, and the blueshifted is due to the approaching side. The hydrogen column density in these clouds is large enough to make the clouds radiation bounded. As a result, the outer parts of the clouds are neutral, and the local NaI atoms can absorb and reemit stellar photons. The blueshifted NaID absorption is due to absorption of stellar photons in the approaching side of the outflow and the redshifted emission is due to absorption of stellar photons in the receding side of the outflow, which are then reemitted isotropically. These photons are not absorbed by the approaching side because of their redshifted central wavelength. This results in the observed P-Cygni-like profile. In this model, the NaID absorption affects both NaID and the ionized line emission (see section \ref{s:emis_decomp}). The roughly constant NaID$_{K}$/H$\alpha$ ratio throughout the FOV suggests that all the outflowing clouds have roughly the same hydrogen column densities, sizes, and neutral-to-ionized gas mass ratios.

The above model is supported by all the observations presented earlier. The neutral and ionized gas phases are believed to be part of the same outflowing clouds due to the observed correspondence in gas kinematics, spatial extent, and constant NaID$_{K}$/H$\alpha$ ratio throughout the FOV (see section \ref{s:connection_between_gas_phases} for details). The full photoionization model presented in section \ref{s:photoionization_models} below is consistent with all these observations.

The suggested double-cone outflow is supported by the fact that we detect redshifted and blueshifted kinematic components in both the ionized and neutral lines (see figures \ref{f:emis_and_abs_fitting_example_neutral_only_with_explanations}, \ref{f:ionized_line_properties_extended}, and \ref{f:emis_and_abs_fitting_example_ionized_only}). We argue that the cones are seen face-on since (i) the maximal positive and negative velocities we observe in the ionized emission lines are similar ($\pm$1\,000 km/sec) and are constant throughout the FOV (see figure \ref{f:ionized_line_properties_extended}), (ii) the NaID emission and absorption coincide throughout the entire FOV, where spaxels with strong emission also show strong absorption and vice versa (see figure \ref{f:naid_properties_all_with_explanations}), and (iii) the central source is classified as an unobscured type I AGN (see section \ref{s:agn_props}). Finally, in section \ref{s:derived_gas_props} we used two different methods to estimate the electron density in the outflow, one which depends on the de-projected distance of the gas and one that does not. By comparing the two methods, we argued that the projected and de-projected locations of the outflow are roughly similar, and are $r \sim $5 kpc. This also suggests that the opening angle of the cone is not small ($\gtrsim 45^{\circ}$). On the other hand, very large opening angles are ruled out since we do not observe lower outflow velocities at the edges of the FOV (see figure \ref{f:ionized_line_properties_extended} and section \ref{s:derived_gas_props}). We therefore suggest that the opening angle of the cone is moderate.

\subsection{Photoionization modeling}\label{s:photoionization_models}

We model the central source of ionizing radiation using standard assumptions about the spectral energy distribution (SED) of AGN. The SED consists of a combination of an optical-UV continuum emitted by an optically-thick geometrically-thin accretion disk, and an additional X-ray power-law source that extends to 50 keV with a photon index of $\Gamma = 1.9$. The normalization of the UV (2500 \AA) to X-ray (2 keV) flux is defined by $\alpha_{OX}$, which we take to be 1.37. We choose an SED with a mean energy of an ionizing photon of 2.56 Ryd. The AGN bolometric luminosity is $1.5\times 10^{44}\,\mathrm{erg/sec}$, which results in a number of ionizing photons of $\mathrm{Q(Lyman) = 1.7 \times 10^{54}}$. The results discussed below depend primarily on the ionization parameter, and are less affected by the specific choices of SED shape and bolometric luminosity. 

The ionization parameter is defined as: $\mathrm{U = Q(Lyman)/4\pi r^2 n_{H} c}$, where $\mathrm{r}$ is the distance of the gas from the central source, $\mathrm{n_{H}}$ is the hydrogen number density, and $\mathrm{c}$ is the speed of light. We estimated the ionization parameter as a function of projected distance from the center of the galaxy, and found ionization parameters in the range $\log{\mathrm{U}}=-3.7$ to $\log{\mathrm{U}}=-3.2$, with the broad emission lines showing a constant $\log{\mathrm{U}}=-3.7$. We therefore focus on three models, with ionization parameters of $\log{\mathrm{U}}=-3.7$, $\log{\mathrm{U}}=-3.4$, and $\log{\mathrm{U}}=-3.2$\footnote{For these low ionization parameters, we do not expect Radiation Pressure Confinement (RPC; e.g., \citealt{dopita02, stern12}) to have a significant effect on the cloud structure, since gas pressure is larger than radiation pressure. Therefore, for the ionization parameters discussed here, constant-pressure models are effectively similar to constant-density models.}. 

Our models consist of optically-thick shells of dusty gas, with ISM-type grains, at a distance of 5 kpc from the central ionizing source. The hydrogen density is determined from the required ionization parameter, and is $100\,\mathrm{cm^{-3}}$, $50\,\mathrm{cm^{-3}}$, and $30\,\mathrm{cm^{-3}}$ respectively. To ensure that the column density is large enough to include neutral Na atoms, we set the total hydrogen column density to be $\mathrm{N_{H}=10^{24}\,cm^{-2}}$. The stellar mass of SDSS J124754.95-033738.6 is estimated to be $\mathrm{M_{*} = 10^{10.8}\, M_{\odot}}$ (e.g., \citealt{chen12}). Therefore, the stellar mass-metallicity relation suggests metallicity which is roughly twice solar (e.g., \citealt{t04}). However, it is unclear whether the metallicity of the outflowing gas should be similar to the metallicity of the stationary ISM in the host galaxy. Therefore, and to allow for a straightforward comparison with previous studies (e.g., \citealt{shih10, rupke15, perna19}), we report results for solar and twice solar metallicity. We note that the main results presented below do not depend strongly on the metallicity.

\begin{figure*}
\includegraphics[width=0.95\textwidth]{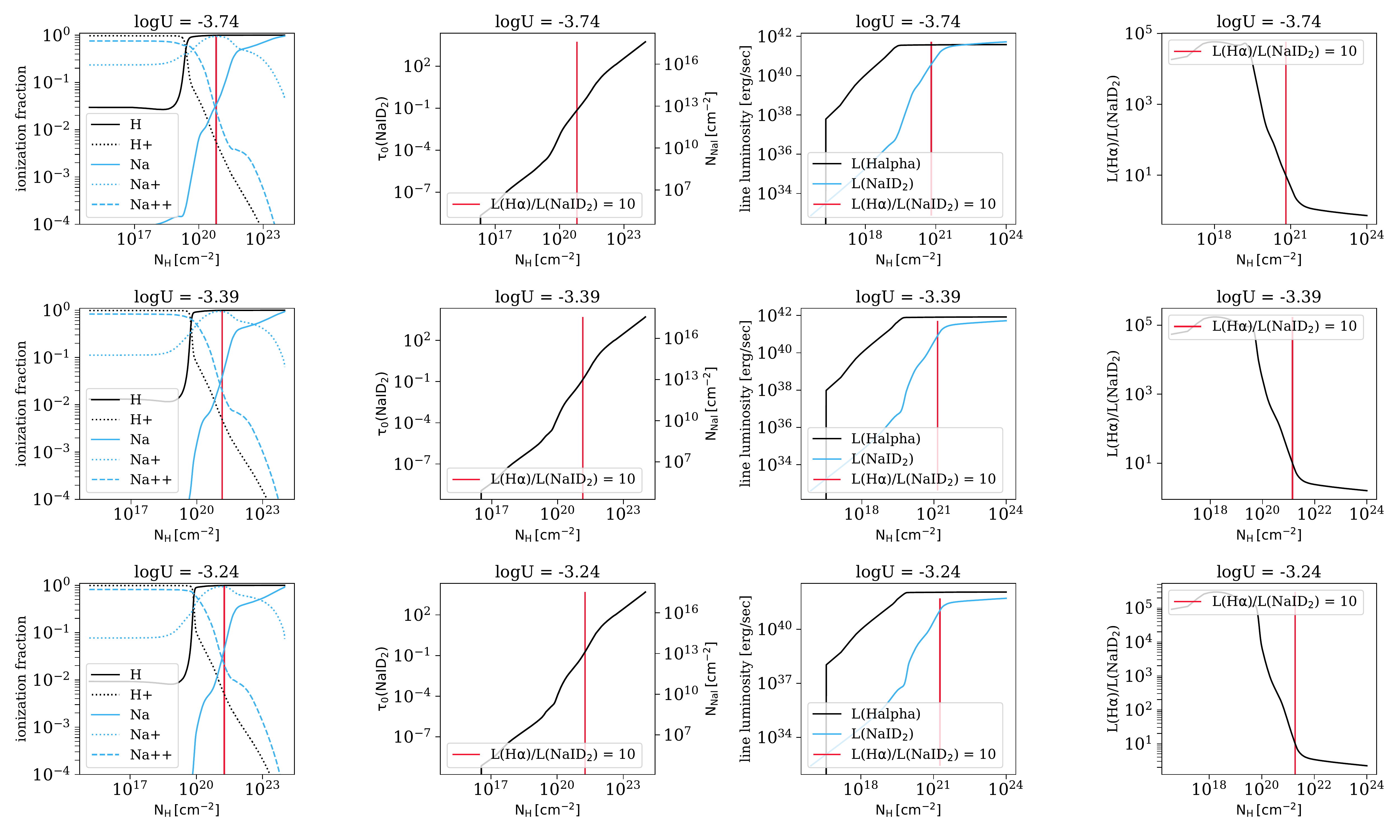}
\caption{\textbf{Photoionization model results.} Photoionization models with solar metallicity, where the rows represent the different ionization parameters, logU=-3.7, logU=-3.4, and logU=-3.2 respectively. The left panels show the ionization fraction of H and Na as a function of the hydrogen column density. The second panels from the left show the optical depth at line center of the NaID$_{K}$ component, and the column density of NaI, as a function of the hydrogen column density in the cloud. The third panels show the expected H$\alpha$ and NaID$_{K}$ line luminosities as a function of hydrogen column density, and the right panels the L(H$\alpha$)/L(NaID$_{K}$) line ratio. We mark the hydrogen column density for which L(H$\alpha$)/L(NaID$_{K}$)=10 using a red line in all the panels. Using these models and the observed L(H$\alpha$)/L(NaID$_{K}$) ratio, we can constrain the hydrogen column density, the Sodium neutral fraction, and the neutral-to-ionized gas mass ratio in the outflow. }\label{f:photoionization_models_selected}
\end{figure*}

In figure \ref{f:photoionization_models_selected} we show the results of the photoionization modeling for solar metallicity, where each row represents a different ionization parameter. In the left panels we show the ionized fractions of H and Na as a function of the hydrogen column density in the cloud. The second panels from the left show the optical depth at line center of the D2 component, $\mathrm{\tau_{0} (NaID_{K})}$, and the column density of neutral Na, $\mathrm{N_{NaI}}$, as a function of the hydrogen column density in the cloud. For solar metallicity, the column density of neutral Na is calculated using:
\begin{equation}\label{eq:9}
	\mathrm{N_{NaI} = N_{H} (1-y) 10^{a + b}},
\end{equation}
where $\mathrm{N_{H}}$ is the hydrogen column density, and $\mathrm{y=1- n(NaI)/n(Na)}$ is the fraction of ionized Na, which is taken from the model. To be consistent with \citet{shih10}, we adopt $\mathrm{a = \log{} \big[ N_{Na}/N_{H} \big]} = -5.69$, which is the solar Na abundance, and $\mathrm{b = \log{} \big[ N_{Na}/N_{H,total} \big] - \log{} \big[ N_{Na}/N{H,gas} \big] = -0.95 }$, which is the ISM depletion of Na. For twice solar metallicity, the righthand side expression in equation \ref{eq:9} is multiplied by two. The optical depth at line center of the $K$ component is then \citep{draine11}:
\begin{equation}\label{eq:10}
	\begin{split}
	& \mathrm{ \tau_{0}(NaID_{K}) = 0.7580 \times } \\ 
	& \times \mathrm{\Big( \frac{N_{Na}}{10^{13}\, cm^{-2}} \Big) \Big(\frac{f_{lu}}{0.4164} \Big) \Big(\frac{\lambda_{lu}}{1215\,\AA}\Big) \Big(\frac{10\,km/sec}{b}\Big)},
	\end{split}
\end{equation}
where $\mathrm{f_{lu}}=0.32$, $\mathrm{\lambda_{lu}=5897}$ \AA, and we take $b=500$ km/sec, which is close to the median velocity width we infer from the broad ionized emission lines and the broad NaID emission and absorption lines.

We calculate the expected H$\alpha$ and NaID emission line luminosities as a function of hydrogen column density. To calculate the H$\alpha$ luminosity, we estimate the emissivity using the expression given in \citet{draine11} for case B recombination, and using the electron temperature and electron density directly from the models. We calculate the expected NaID luminosity using: $\mathrm{L(NaID_{K}) = EW_{abs}(NaID_{K}) \times L_{\lambda; stars}(\lambda = 5987\,\AA)}$, where $\mathrm{EW_{abs}(NaID_{K})}$ is the EW of the NaID$_{K}$ absorption, which we calculate using the optical depth at line center ($\mathrm{\tau_{0}(NaID_{K})}$, equation \ref{eq:10}), and $\mathrm{L_{\lambda; stars}(\lambda = 5987\,\AA)}$ is the stellar continuum emission at $\lambda = 5897$\AA, which we take directly from observations. The dust-corrected stellar continuum emission at $\mathrm{\lambda = 5897}$\AA\, in the central spaxel is $3\times 10^{43}\,\mathrm{erg/(sec\, cm^{2})}$, and the integrated stellar continuum emission at $\mathrm{\lambda = 5897}$\AA\, is $2\times 10^{44}\,\mathrm{erg/(sec\, cm^{2})}$. These are one and two orders of magnitude larger than the expected AGN continuum emission at the same wavelength. Therefore, the NaID emission is powered by the stellar continuum emission and not by the AGN. Thus, according to our suggested model, while the AGN is responsible for the ionization structure in the cloud, the NaID line luminosity is powered by the stellar continuum.

In the third panels of figure \ref{f:photoionization_models_selected} we show the H$\alpha$ and NaID$_{K}$ line luminosities, for a unit covering factor, as a function of the hydrogen column density. The H$\alpha$ luminosity increases with the ionized column all the way to the ionization front, at $\mathrm{N_H}$ of a few $10^{19}\, \mathrm{cm^{-2}}$,  beyond which very few line photons are created. The NaID$_{K}$ line luminosity increases as a function of depth into the cloud, with the steepest rise beyond the ionization front. The saturation of NaID$_{K}$ luminosity occurs when the absorption line transitions from the optically-thin regime to the flat part of the curve of growth. 

In the right panels of figure \ref{f:photoionization_models_selected} we show the L(H$\alpha$)/L(NaID$_{K}$) line ratio as a function of the hydrogen column density, where we mark the observed ratio of L(H$\alpha$)/L(NaID$_{K}$)=10 using a red line. This sets the physical size of the cloud, the total dust column density and extinction, the ionized fraction of Na atoms, and the optical depth and EW of the NaID$_{K}$ absorption. In table \ref{tab:natbib} we list several of the resulting properties of the clouds, for solar and twice solar metallicities. Interestingly, these models reproduce, to within a factor of 2--3, all the observed properties of the NaID$_{K}$ and the dust in our system. In particular, for the same ionization parameter to that observed, the resulting NaID$_{K}$ absorption optical depth and EW are roughly similar to those we measure. More importantly, the derived dust reddening from the models is roughly consistent with the estimated reddening from the \emph{ionized emission lines}. This correspondence suggests that the proposed model, where the ionized and neutral phases are part of the same outflowing clouds, is consistent with the observations.

One can see that the main difference between the solar and twice solar metallicity models is in the hydrogen column density for which L(H$\alpha$)/L(NaID$_{K}$)=10, where models with solar metallicity require twice as much $\mathrm{N_{H}}$ to reach the requirement. This also sets the different neutral-to-ionized gas mass ratio, where models with solar metallicity require twice as much neutral-to-ionized gas mass ratio. 

\begin{table*}
 \caption{Results of the photoionization models with solar and twice solar metallicities. The columns from left to right are: (1) the ionization parameter of the ionized gas, (2) gas metallicity, (3) hydrogen column density at the ionization front, (4) hydrogen column density at L(H$\alpha$)/L(NaID$_{K}$)=10, (5) the neutral-to-ionized gas mass ratio at L(H$\alpha$)/L(NaID$_{K}$)=10, (6) the inferred dust reddening at L(H$\alpha$)/L(NaID$_{K}$)=10, (7) the inferred optical depth at line center at L(H$\alpha$)/L(NaID$_{K}$)=10, (8) the inferred EW at L(H$\alpha$)/L(NaID$_{K}$)=10, and (9) the neutral fraction of Sodium at L(H$\alpha$)/L(NaID$_{K}$)=10. }\label{tab:natbib}

\begin{tabular}{ccccccccc}
\hline
(1) & (2) & (3) & (4) & (5) & (6) & (7) & (8) & (9) \\
$\mathrm{log{U}}$ & $Z$ & $\mathrm{N_{H}}$(H$_{0}$=H$^{+}$) & $\mathrm{N_{H}(L_{H\alpha}/L_{NaID_{K}}=10)}$ & $\mathrm{\frac{M_{neutral}}{M_{ionized}}}$ & $\mathrm{E}_{B-V}$ & $\tau_{0, K}$ & EW(NaID$_{K}$) & $\mathrm{(1 - y) }$ \\
   & [$\mathrm{Z_{\odot}}$] & [$\mathrm{cm^{-2}}$] & [$\mathrm{cm^{-2}}$] & & [mag] &  & [\AA\,] &  \\ 
  \hline
  -3.7 & 1 & $2.5\times10^{19}$ & $9.0 \times 10^{20}$ & 36 & 0.16 & 0.03 & 0.87 & 0.05 \\ 
  -3.7 & 2 & $2.1\times10^{19}$ & $4.0 \times 10^{20}$ & 19 & 0.14 & 0.02 & 0.87 & 0.05 \\ 
  \hline
  -3.4 & 1 & $5.4\times10^{19}$ & $1.8 \times 10^{21}$ & 34 & 0.32 & 0.05 & 2.0 & 0.05 \\ 
  -3.4 & 2 & $4.4\times10^{19}$ & $8.2 \times 10^{20}$ & 19 & 0.29 & 0.05 & 1.9 & 0.06 \\   
 \hline
 -3.2 & 1 & $7.8\times10^{19}$ & $2.3 \times 10^{21}$ & 32 & 0.42 & 0.06 & 2.7 & 0.07 \\ 
 -3.2 & 2 & $6.1\times10^{19}$ & $1.1 \times 10^{21}$ & 18 & 0.38 & 0.06 & 2.6 & 0.07 \\
  \hline
\end{tabular}
\end{table*}

Having a consistent photoionization model, we can use the model to put constrains on additional properties of the clouds. First, we find that the mass of the outflowing neutral gas is 20--40 times larger than the outflowing ionized gas, depending on the gas metallicity. This is somewhat larger than the neutral-to-ionized gas mass ratio we suggested for typical ionized outflows in type II AGN \citep{baron19b}. Secondly, we can use the observed narrow and broad L(H$\alpha$) to constrain the covering factor of the outflowing gas. For the assumed AGN bolometric luminosity and for a covering factor of 1, the reddening-corrected luminosity is L(H$\alpha$)$=2.25\times 10^{42}\,\mathrm{erg/sec}$. We measure dust-corrected L(H$\alpha$)$=1.57\times 10^{42}\,\mathrm{erg/sec}$ and L(H$\alpha$)$=8\times 10^{41}\,\mathrm{erg/sec}$ for the broad and narrow components respectively, suggesting a covering factor of 0.7 for the broad outflowing gas and 0.35 for the narrow component. The large covering factor we find for the outflow is in line with the large covering factor required in the NaID fitting. The covering factor of the narrow component, 0.35, is higher than the covering factor of typical NLR in AGN (e.g., \citealt{baskin05, mor09}). This reinforces our suggestion that a large fraction of the narrow component is not stationary NLR, but rather part of the outflow. 

We can also estimate the size of the outflowing clouds. Assuming a distance of 5 kpc for the outflowing clouds and a solar (twice solar) metallicity, the size of the clouds is 3 pc, 12 pc, and 25 pc (1.5 pc, 6 pc, 12 pc) for ionization parameters of $\log{\mathrm{U}}=-3.7$, $\log{\mathrm{U}}=-3.4$, and $\log{\mathrm{U}}=-3.2$ respectively. Finally, we can use the photoionization models and the required L(H$\alpha$)/L(NaID$_{K}$)=10 to constrain the neutral fraction of Sodium in the cloud, $(1 - y)$. In earlier works of this type, the neutral fraction was assumed to be $(1 - y)=0.1$, which is similar to the value measured in the Milky-Way (e.g., \citealt{rupke05a, rupke05b, shih10, rupke15, rupke17}). We measure $(1 - y) \sim 0.05$. To the best of our knowledge, this is the first time that $(1- y)$ is estimated, rather than assumed, in an outflow. In section \ref{s:mass_and_energy} below we use our best fit value of $y$ to estimate the mass and energetics of the observed outflows.

\textbf{ In summary,} the photoionization models account for both the ionized and the neutral gas phases. We find consistent L(H$\alpha$)/L(NaID) emission line ratios, consistent ionization parameters and ionized line ratios, consistent NaID absorption optical depths and EWs, and consistent dust reddening, to those observed. In addition, we used the photoionization model and the observed L(H$\alpha$)/L(NaID) line ratio to constrain, for the first time, the size, the Sodium neutral fraction, and the neutral-to-ionized gas mass ratio in the outflowing clouds. The size of each of the outflowing clouds is 1.5--3 pc, the Sodium neutral fraction is 0.05, and the neutral-to-ionized gas mass ratio is 20--40. To the best of our knowledge, this is one of first cases in which all these properties were estimated in a single outflow. 

\subsection{Mass outflow rate and energetics}\label{s:mass_and_energy}

Our model allows us to estimate the outflowing gas mass and its kinetic energy. For the ionized gas we use the expressions given in section 5 in \citet{baron19b}. These expressions require the dust-corrected H$\alpha$ (or [OIII]) line luminosity\footnote{Note that the broad H$\alpha$ and [OIII] line luminosities are corrected for dust reddening using the extinction derived towards the \emph{broad} lines, not the narrow lines.}, the electron density in the outflow, the gas emissivity, the location of the outflow, and the outflow velocity. It is important to note that to estimate the ionized gas mass using recombination or forbidden lines, we do not need to have knowledge of the outflow geometry (such as covering factor, filling factor, and the entire 3D spatial distribution of the gas), since the measured line luminosity, which is used to estimate the mass, already encompasses this information (see e.g., \citealt{cano12}). This is contrary to the estimate of neutral gas mass using absorption lines, where one must assume a geometry, and have a knowledge of the covering factor and filling factor of the wind.

To estimate the mass and energetics of the ionized outflows, we use the measured broad H$\alpha$ luminosities and the electron densities derived using the ionization parameter method. We assume the emissivities calculated in section 4.3 in \citet{baron19b}, and assume that the outflow location is 5 kpc from the central BH. We take the velocity of the outflow to be the maximal positive velocity, $\Delta v + 2\sigma$ (see bottom row of figure \ref{f:ionized_line_properties_extended}). We find an ionized outflowing gas mass of $\mathrm{M_{ion} = 8 \times 10^6 \, M_{\odot}}$, a mass outflow rate of $\mathrm{\dot{M}_{ion} = 1 \, M_{\odot}/yr}$, kinetic power of $\mathrm{\dot{E}_{ion} = 1.4 \times 10^{41}\, erg/sec}$, and a kinetic coupling efficiency of $\mathrm{\epsilon_{ion} = 10^{-3}}$. 

Next, we estimate the \emph{neutral} gas mass, mass outflow rate, kinetic power, and kinetic coupling efficiency, using the expressions given in section 5.3 in \citet{shih10}, pertaining to the time-averaged thin-shell model. It has been widely used to estimate the outflowing gas mass from absorption line spectroscopy (e.g., \citealt{rupke05a, rupke05b, rupke05c, shih10, cazzoli16, rupke17}). This model requires the hydrogen column density in the outflow, the gas covering factor, the location of the outflow, and the outflow velocity. The hydrogen column density is estimated using equations \ref{eq:9} and \ref{eq:10}, combined with the measured properties of the NaID \emph{absorption}. We assume twice solar metallicity, and use the measured velocity width $\mathrm{b = \sqrt{2} \sigma_{broad}}$, where $\mathrm{\sigma_{broad}}$ is the velocity dispersion in the broad emission and absorption lines (see figure \ref{f:ionized_line_properties_extended}). Since we estimate the mass and energetics of the neutral gas with the NaID absorption, we take the covering factor to be 0.7, similar to what we found in our photoionization model. Similarly to the ionized outflow, we assume that the outflow location is 5 kpc and take the velocity to be the maximal positive velocity, $\Delta v + 2\sigma$.

We find a neutral outflowing gas mass of $\mathrm{M_{neutral} = 1.8 \times 10^8 \, M_{\odot}}$, a mass outflow rate of $\mathrm{\dot{M}_{neutral} = 26 \, M_{\odot}/yr}$, kinetic power of $\mathrm{\dot{E}_{neutral} = 4.4 \times 10^{42}\, erg/sec}$, and a kinetic coupling efficiency of $\mathrm{\epsilon_{neutral} = 0.03}$. The measured neutral-to-ionized gas mass ratio is 23, and the neutral-to-ionized mass outflow rate ratio is 26. These values are very close to the neutral-to-ionized gas mass ratio we calculated using the photoionization models in section \ref{s:photoionization_models}. This consistency is not a trivial consequence of our assumptions. To estimate the ionized and neutral gas mass and outflow rate, we used the dust-corrected H$\alpha$ luminosity and the NaID absorption line optical depth respectively. For the former, it is not required to assume a geometry, while for the latter we assumed the time-averaged thin-shell model with a covering factor of 0.7. Furthermore, the expressions for the ionized and neutral gas mass depend \emph{differently} on $r$, the outflow location. Therefore, the consistent neutral-to-ionized mass ratio justify the assumptions about the geometry and outflow location. 

These estimates pertain only to the broad emission lines.

\subsection{Evidence that the narrow emission lines are part of the outflow}\label{s:origin_of_narrow_lines}

A standard assumption in studies of ionized outflows in AGN is that the emission lines can be decomposed into two components: a narrow core which originates from stationary gas in the galaxy, and a broad component which is emitted by fast-moving outflowing clouds (see e.g., \citealt{mullaney13, karouzos16a, baron17b, rupke17, baron19b}). This is obviously a simplified assumption. For example, in \citet{baron18} we detected an outflow in an almost edge-on configuration, where the emission lines that were emitted by the outflow were narrow due to projection effects. More generally, the 3D biconical outflow models of \citet{bae16} show that emission lines that originate in the outflow present complex line kinematics, and depending on the geometry, can exhibit both narrower and broader components. Thus, a decomposition into narrow and broad kinematic components, where the former represents the stationary gas and the latter represents the outflow, may underestimate the total mass outflow rate.

Regardless of the question of whether typical NLRs host stationary or outflowing gas\footnote{Here, "typical NLR" is defined as the region that emits the narrow emission lines in AGN. This region is not necessarily stationary, and could host an outflow component.}, we find several differences between the properties of the narrow line-emitting gas in our source and the properties of many other well studied NLRs. First, we detect narrow NaID emission with the same kinematics as the narrow ionized emission lines, suggesting that both of these are emitted by the same clouds. This is very different from typical NLRs, where narrow NaID \emph{emission} was not detected so far. Second, we found in section \ref{s:photoionization_models} that the covering factor of the narrow line-emitting gas is 0.35, which is significantly higher than the covering factor observed in typical NLRs (e.g., \citealt{baskin05, mor09}). Moreover, if we use the dust corrected narrow optical line luminosities to estimate the AGN bolometric luminosity (see \citealt{netzer09, netzer19}), we find a bolometric luminosity which is $\sim$5 times higher than the bolometric luminosity we infer from the 2--10 keV luminosity (section \ref{s:agn_props}). These suggest that the narrow line luminosities in this source are significantly higher than those expected from a typical NLR of an AGN with the same bolometric luminosity. Therefore, the narrow lines in our source are not emitted by a typical NLR.

The most direct observational evidence that the narrow emission originates in the outflow is the detection of a narrow redshifted NaID emission line. Since the emitted NaI radiation is only due to the scattering of the absorbed stellar continuum, this emission cannot be explained with a stationary NLR-like gas (unless a special geometry is invoked) and requires an outflow. Since we argue that the neutral and ionized emission lines are emitted by the same clouds, this suggests that the narrow ionized emission lines are emitted by the outflow as well.

To estimate the exact contribution of the outflow to the narrow emission lines, it is necessary to perform full dynamical modeling of the system, which is beyond the scope of this paper. However, we can estimate the mass and energetics of the observed outflows in the case where most of the narrow lines are emitted by the outflow. The dust-corrected narrow line luminosities are roughly half of the broad line luminosities. Combining the two kinematic components we find that the outflowing gas mass is $\mathrm{M_{ion} = 1.2 \times 10^7 \, M_{\odot}}$, the mass outflow rate is $\mathrm{\dot{M}_{ion} = 1.5 \, M_{\odot}/yr}$, the kinetic power is $\mathrm{\dot{E}_{ion} = 2.1 \times 10^{41}\, erg/sec}$, and the kinetic coupling efficiency is $\mathrm{\epsilon_{ion} = 1.5 \times 10^{-3}}$. For the neutral gas, the gas mass is $\mathrm{M_{neutral} = 2.7 \times 10^8 \, M_{\odot}}$, the mass outflow rate is $\mathrm{\dot{M}_{neutral} = 39 \, M_{\odot}/yr}$, the kinetic power is $\mathrm{\dot{E}_{neutral} = 6.6 \times 10^{42}\, erg/sec}$ and the kinetic coupling efficiency is $\mathrm{\epsilon_{neutral} = 0.045}$.

\section{Summary and conclusions}\label{s:concs} 

This work is part of a long-term project to map and analyze multi-phase AGN-driven winds at the specific evolutionary stage of post starburst E+A galaxies. In this paper, we present new MUSE/VLT observations of SDSS J124754.95-033738.6, at z=0.09. The spatially-resolved spectroscopy presented here allowed us to study the stellar and the neutral and ionized gas properties throughout the entire FOV. Our results can be summarized as follows:
\begin{enumerate}
\item SDSS J124754.95-033738.6 is a spiral galaxy with spiral arms extending to distances of $\sim$20 kpc. Its optical spectrum suggests a post starburst E+A galaxy. While its stars are showing typical stellar rotation, we detect spatially and kinematically disturbed gas throughout the entire FOV, suggestive of a minor merger that probably triggered the recent burst.
\item Using stellar population synthesis modeling, we find that the system had two SF episodes, with a recent short episode that started 70 Myrs ago and ended 30 Myrs ago, and an older long episode that started 14 Gyrs ago and ended 6 Gyrs ago. The total stellar mass of the system is $\mathrm{M_{*} = 10^{10.8}\,M_{\odot}}$, where 2\% of the mass was formed in the recent burst. The estimated peak SFR during the recent burst is about 150 $\mathrm{M_{\odot}}$/yr. The ionized emission lines in the central spaxels suggest residual SF that is obscured by dust, with estimated SFR of $2.2\,\mathrm{M_{\odot}/yr}$. This SFR places the system roughly 0.4 dex below the SF main sequence.
\item We detect two kinematic components that trace ionized gas, both of which are ionized by the central AGN. Our analysis of the broad kinematic component suggests a fast-moving, $v \sim 1\,000$ km/sec, ionized outflow, in a face on double-cone configuration with a moderate opening angle of about $45^{\circ}$. We further suggest that most of the narrow line emission originates in the outflow as well, where the total emission line profile is shaped by projection effects and dust reddening.
\item We detect NaID \emph{emission} and absorption throughout a large fraction of the FOV. We detect narrow and broad kinematic components in emission, and broad kinematic components in absorption, suggesting a fast-moving, $v \sim 1\,000$ km/sec, neutral outflow. This is the third reported case of resolved NaID emission in an outflow. 
\item We find a remarkable similarity between the kinematics and spatial extents of the ionized and the neutral gas. Furthermore, we find a constant L(H$\alpha$)/L(NaID) luminosity ratio throughout the entire FOV, suggesting that the H$\alpha$-emitting and NaID-emitting gas are part of the \emph{same} clouds.
\item We propose a model in which the ionized line emission and the neutral line emission and absorption originate in the same clouds, which are embedded in a fast-moving double-cone outflow. These outflowing clouds are exposed to the AGN radiation, which results in ionized gas which emits the observed emission lines. The back of the cloud, which is neutral, is exposed to the stellar radiation field and produces the observed NaID emission and absorption lines. This model naturally explains the observed kinematic connection between the two phases, and the detection of both NaID emission and absorption.
\item We present photoionization models of the outflowing clouds, which account for both the ionized and the neutral gas phases. The models successfully reproduce various observed properties of the emission and absorption lines. In particular, we find consistent L(H$\alpha$)/L(NaID) emission line ratios, consistent ionization parameters and ionized line ratios, consistent NaID absorption optical depths and EWs, and consistent dust reddening values, to those we observe. 
\item Using the photoionization model, we are able to estimate the properties of the outflowing clouds. The predicted size of the clouds is 1.5--3 pc, and the neutral-to-ionized gas mass ratio is 20--40. To the best of our knowledge, this is one of the first cases in which these properties are directly measured in an outflow.
\item The best-fitting photoionization model allowed us to constrain, for the first time, the neutral fraction of Sodium atoms within the clouds, $(1- y) = 0.05$. The neutral fraction is of particular importance in the context of neutral outflows, since the estimated outflow mass and mass outflow rate depend on it.
\item We estimate the mass and energetics of the ionized and neutral outflows in our system. For the ionized gas, the outflowing gas mass is $\mathrm{M_{ion} = 1.2 \times 10^7 \, M_{\odot}}$, the mass outflow rate is $\mathrm{\dot{M}_{ion} = 1.5 \, M_{\odot}/yr}$, the kinetic power is $\mathrm{\dot{E}_{ion} = 2.1 \times 10^{41}\, erg/sec}$, and the kinetic coupling efficiency is $\mathrm{\epsilon_{ion} = 1.5 \times 10^{-3}}$. For the neutral gas, the gas mass is $\mathrm{M_{neutral} = 2.7 \times 10^8 \, M_{\odot}}$, the mass outflow rate is $\mathrm{\dot{M}_{neutral} = 39 \, M_{\odot}/yr}$, the kinetic power is $\mathrm{\dot{E}_{neutral} = 6.6 \times 10^{42}\, erg/sec}$ and the kinetic coupling efficiency is $\mathrm{\epsilon_{neutral} = 0.045}$. 
\item We measure neutral-to-ionized gas mass ratio of 23, and neutral-to-ionized mass outflow rate ratio of 26, both very close to the ratio we calculated using our photoionization models. This consistency is not a trivial consequence of our assumptions. It confirms the accuracy of our estimates (outflow extent, dust reddening, covering factor, etc) are our assumed thin-shell geometry. 
\end{enumerate}

The E+A galaxy SDSS J124754.95-033738.6 shows one of the most direct connections between the neutral and ionized outflow phases, and thus allowed us to estimate several previously-unknown properties of the outflowing clouds. It remains to be seen whether additional post starburst E+A galaxies show similar connections between the different outflowing phases. We are currently involved in a detailed analysis of similar objects and results will be reported in forthcoming publications.

\section*{Acknowledgments}
We thank the referee, R. Maiolino, for useful comments and suggestions that helped improve this manuscript.
We thank S. Cazzoli, M. Perna, T. Shimizu, and H. Yesuf for useful discussions regarding different aspects presented in this manuscript. 
D. Baron is supported by the Adams Fellowship Program of the Israel Academy of Sciences and Humanities.
This research made use of {\sc Astropy}\footnote{http://www.astropy.org}, a community-developed core Python package for Astronomy \citep{astropy2013, astropy2018}.

This work made use of SDSS-III\footnote{www.sdss3.org} data. Funding for SDSS-III has been provided by the Alfred P. Sloan Foundation, the Participating Institutions, the National Science Foundation, and the U.S. Department of Energy Office of Science. SDSS-III is managed by the Astrophysical Research Consortium for the Participating Institutions of the SDSS-III Collaboration including the University of Arizona, the Brazilian Participation Group, Brookhaven National Laboratory, Carnegie Mellon University, University of Florida, the French Participation Group, the German Participation Group, Harvard University, the Instituto de Astrofisica de Canarias, the Michigan State/Notre Dame/JINA Participation Group, Johns Hopkins University, Lawrence Berkeley National Laboratory, Max Planck Institute for Astrophysics, Max Planck Institute for Extraterrestrial Physics, New Mexico State University, New York University, Ohio State University, Pennsylvania State University, University of Portsmouth, Princeton University, the Spanish Participation Group, University of Tokyo, University of Utah, Vanderbilt University, University of Virginia, University of Washington, and Yale University. 

\bibliographystyle{mn2e}
\bibliography{ref_MUSE_1}

\clearpage

\appendix

\section{Integrated line fluxes}\label{a:appendix}

\begin{figure*}
\includegraphics[width=0.95\textwidth]{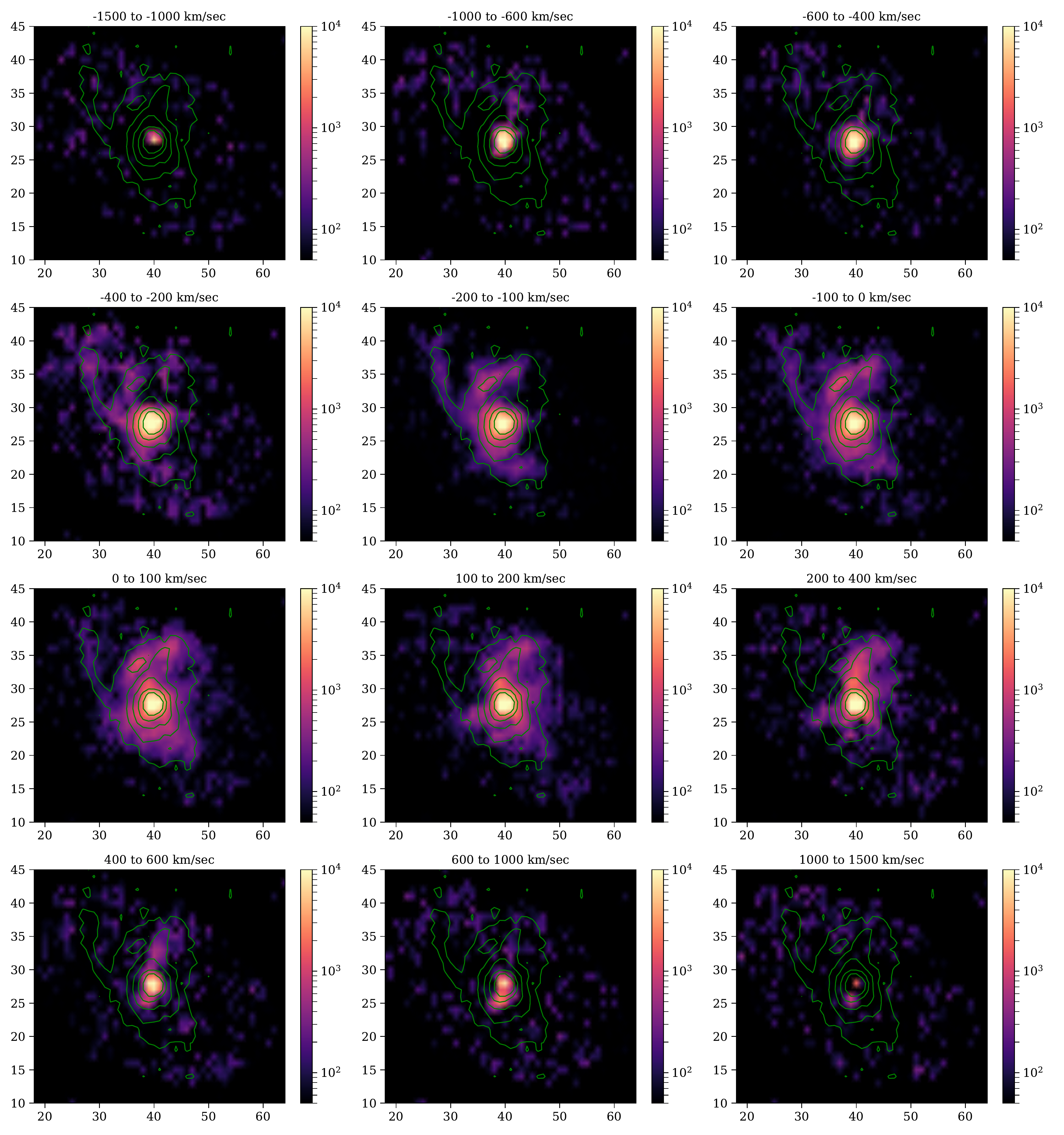}
\caption{Integrated H$\beta$ and [OIII]~$\lambda \lambda$ 4959,5007\AA\, line flux for different velocity channels throughout the FOV, with a logarithmic color-coding and physical units of $10^{-20}\, \mathrm{erg/(sec\, cm^{2} \AA)}$. The green contours represent steps of 0.5 dex in integrated gas emission, traced by the H$\alpha$+[NII] lines (see figure \ref{f:ionized_gas_emission}).
}\label{f:oiii_and_hbeta_chanel_maps}
\end{figure*}

\begin{figure*}
\includegraphics[width=0.9\textwidth]{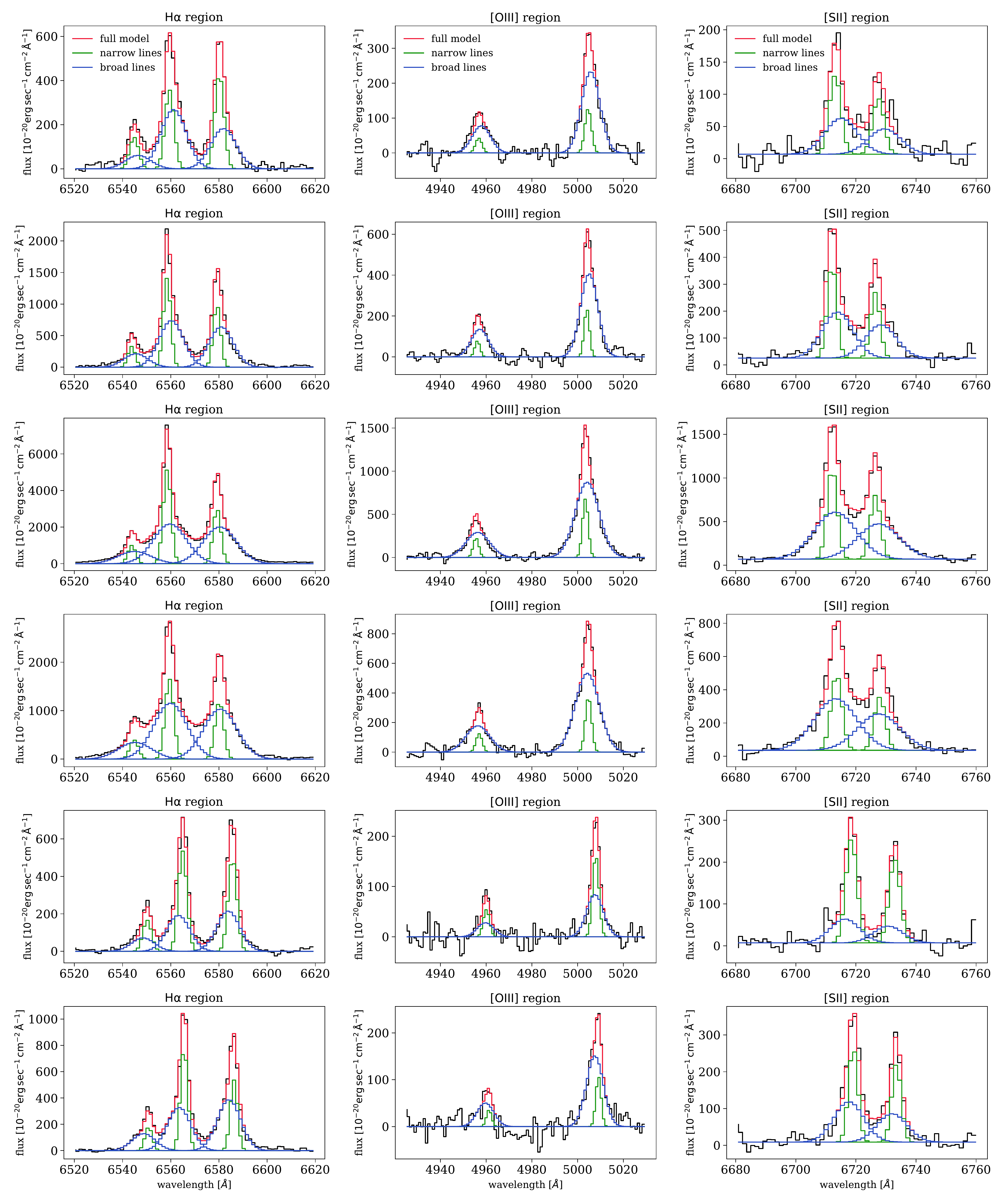}
\caption{\textbf{Six representative spectra from different spaxels and their best-fitting models for the ionized emission lines.} Each row represents the spectrum of a different spaxel. The leftmost panels show the H$\alpha$+[NII] emission line complexes, which are used first in the emission line decomposition. The full model is marked with a red line, and the narrow and broad kinematic components are marked with green and blue lines respectively. The two middle panels show the [OIII] and [SII] emission line regions, with similar color-coding.
}\label{f:emis_and_abs_fitting_example_ionized_only}
\end{figure*}

\end{document}